\documentclass[useAMS,usenatbib]{mn2e}
              \usepackage{times}
              \usepackage{graphicx}
              \title[Galactic restrictions on iron production]
       {Galactic restrictions on iron production by various Types of supernovae}
       
              \author[Acharova et. al.]
              {I.A. Acharova, $^{1}$\thanks{E-mail:
              iaacharova@sfedu.ru (IAA); unmishurov@sfedu.ru (YuNM);\,\,\, val@deneb1.odessa.ua (VVK)}
              Yu.N. Mishurov, $^{1,2}$ and V.V. Kovtyukh $^{3,4}$\\
              $^{1}$ Department of Physics of Cosmos, Southern Federal University, 
              5 Zorge, Rostov-on-Don, 344090, Russia \\
       $^{2}$ Special Astrophysical Observatory of the Russian Academy of Sciences, 
              N.Arkhyz, Karachaevo-Cherkessia, Russia\\
              $^{3}$ Astronomical Observatory of Odessa National University, Shevchenko Park, 65014 Odessa, Ukraine\\
       $^4$ Isaac Newton Institute of Chile, Odessa Branch,
       Ukraine\\
                }

\begin{document}

              \date{Accepted 2011 xxxx. Received 2011 xxxx; in original form 2011 xxxx}
              
              \maketitle
              
              \pagerange{\pageref{firstpage}--\pageref{lastpage}} \pubyear{2011}
              
              \label{firstpage}
              
              \begin{abstract}
       We propose a statistical method for decomposition of contributions to iron production from various sources: supernovae Type II and the subpopulations of supernovae Type Ia -- prompt (their progenitors are short-lived stars of ages less then $\sim$100 Myr) and tardy (whose progenitors are long-lived stars of ages $>$100 Myr). To do that, we develop a theory of oxygen and iron synthesis which takes into account the influence of spiral arms on amount of the above elements synthesized by both the supernovae Type II and prompt supernovae Ia. In the framework of the theory we processed statistically the new more precise observational data on Cepheids abundances, which, as it is well known, demonstrate nontrivial radial distributions of oxygen and iron in the Galactic disc with bends in the gradients. In our opinion, such fine structure in the distribution of the elements along the Galactic disc enables to decompose unambiguously the amount of iron into 3 components produced by the above 3 sources of it. Besides, by means of our statistical methods we solve this task without of any preliminary suppositions about the ratio among the portions of iron synthesized by the above sources. 
       
       The total mass supplied to the Galactic disc during its life by all Types of SNe happens to be $\sim (4.0 \pm 0.4)\cdot 10^7$ M$_{\odot}$, the mass of iron occurs in the present ISM is $\sim (1.20 \pm 0.05)\cdot 10^7$ M$_{\odot}$, i.e., about 2/3 of iron is contained in stars and stellar remnants.

       The relative portion of iron synthesized by tardy supernovae Ia for the life-time of the Galaxy is $\sim$35 per cent (in the present ISM this portion is $\sim$50 per cent). Correspondingly, the total portion of iron supplied to the disc by supernovae Type II and prompt supernovae Ia is $\sim$65 per cent (in the present ISM this portion is $\sim$50 per cent). 
       The above result slightly depends on the adopted mass of oxygen and iron synthesized during one explosion of supernovae and the shape (bimodal or smooth) of the so-called Delay Time Distribution function.

       The portions of iron mass distributed between the short-lived supernovae are usually as follows: depending on the ejected masses of oxygen or iron during one supernovae Type II event the relative portion of iron, supplied to the Galactic disc for its age, varies in the range 12 - 32 per cent (in the present ISM 9-25 per cent); the portion supplied by prompt supernovae Ia to the Galactic disc is 33 - 53 per cent (in ISM 26 - 42 per cent).
       
       Our method also confirm that the bend in the observed slope of oxygen radial distribution and the minimum in [O/Fe] at $\sim$7 kpc form in the vicinity of location of the corotation resonance.

              \end{abstract}
                     
              \begin {keywords}
              Galaxy: fundamental parameters -- Galaxy: abundances -- ISM: abundances -- galaxies: spiral -- galaxies: star formation -- ({\it stars:}) supernovae general.
              \end{keywords}

                     
              \section{Introduction}

       In the present paper, we extend the statistical method, proposed by Acharova et al. (2011) for analysis of the radial distribution of oxygen in the Galaxy Milky Way, to explain the nontrivial distribution of iron along the Galactic disc, revealed in a series papers by Andrievsky et al. (2002 a,b,c) and Luck et al. (2003; 2006; 2011). This problem is of great importance not only for the chemical evolution of the Galactic disc and the history of star formation, but also for the search of independent restrictions on the models of supernovae, especially SNe Type Ia (SNe Ia) whose outstanding role in discovery of accelerated expansion of the Universe is well known (Riess et al. 1998; Perlmutter et al. 1999).
       
       According to the cited papers of Andrievsky, Luck and their collaborators the spectroscopic study of heavy element abundances in Cepheids demonstrate that, in the Milky Way, the radial distributions both oxygen and iron are to be described by a multi-slope function rather than by a linear one. For instance, the distribution of oxygen along the Galactic disc is characterized by a steep gradient in the inner part of the disc for Galactocentric radii 5 $\le r \le$ 7 kpc and a plateau-like distribution for $r >$ 7 kpc and up to 10 kpc (for the solar Galactocentric distance $r_0$ the value 7.9 kpc is adopted), so that at $r \sim$ 7 kpc there is an inflection in the slope of the distribution.

       This fine structure of the radial distribution of oxygen was first explained by Acharova et al. (2005 a,b). For this, they took into account the influence of spiral arms since oxygen is mainly synthesized during explosions of SNe Type II (SNe II) which are strongly concentrated in spiral arms. As it was shown in the last papers the combined effect of the corotation resonance and turbulent diffusion results in formation of the radial distribution of oxygen in the Galactic disc with the bend in the slope.

       But some time ago it was difficult to explain the similar behavior of iron along the Galactic radius by means of the same mechanism since it is generally agreed that $\sim$~70 \% of iron is synthesized during SNe Ia explosions and only $\sim$~30 \% is produced by SNe II (Matteucci 2004). The point is that the progenitors of SNe Ia were thought to be of ages of the order of several billion years. Before outbursts, after such a long period of time they have to be dispersed over a very large portion of the Galactic disc (Mishurov \& Acharova 2011). Hence, if all precursors of SNe Ia would be old stars, they did not keep in their memory that they were born in spiral arms. So, we could not expect any noticeable influence of spiral arms on the radial distribution of iron unless to increase significantly the output of iron per one SNe II event which would entail serious consequences in the theory of pre SNe II evolution.

       The opportunity to solve simultaneously the problem of formation of the above fine structure in the radial distribution both oxygen and iron was offered when it has become evident that there exist two subpopulations of SNe Ia precursors~-- short-lived and long-lived. They were called `prompt' and `tardy', respectively (see, e.g., Mannucci et al. 2005; 2006; Aubourg et al. 2008; Brandt et al. 2010; Maoz et al. 2011; Li et al. 2011). Acharova et al. (2010) incorporated the results of Mannucci et al. (2006) and Matteucci et al. (2006) in their theory and explained the formation of the above fine structure of iron radial distribution in the Galactic disc. Besides, the discovery of the 2 subpopulations of SNe Ia enables to understand their concentration in spiral arms first revealed by Bartunov et al. (1994). 
       \footnote {By the way, they also suggested that, in spiral and elliptical galaxies, SNe Ia appear to have different origin.}

       Nevertheless, several questions arise. First, to decompose the contributions of the above 3 sources to iron synthesis, Acharova et al. (2010) assume that SNe II and the two subpopulations of SNe Ia supply to the present ISM approximately equal portions of iron. In a certain sense, it is a rather arbitrary supposition. However, the extension of the statistical method proposed by Acharova et al. (2011) enables to estimate the contributions to iron production from various Types of SNe independently of any preliminary suppositions. Second, the estimations for ages of prompt progenitors of SNe Ia (denote them as SNe Ia-P) vary from $\sim$~100 Myr and up to $\sim$~400 Myr or even more, depending on used methods of data processing and observational material. Can the radial distribution of iron along the Galactic disc apply restriction on the age of SNe Ia-P precursors? Third, for the so-called `DTD' (Delay Time Distribution) function
       \footnote {DTD function is the probability distribution of the time period between the SNe progenitor birth and explosion.}
       were proposed two different approximations: a bimodal-like (e.g., Mannucci et al. 2006; Matteucci et al., 2006) and smooth one peaked at early times (e.g., Maoz et al. 2010). How do such different representations for DTD function result in metallicity distribution along the Galactic disc and the amount of iron synthesized by various Types of SNe? The answer this question, perhaps, will impose additional constraints on the models of SNe Ia which have not been fully built, as yet.

       At the time, the discussed discovery poses a problem for cosmology, as well. Indeed, if the population of SNe Ia is inhomogeneous, can we consider them as standard candles (Maeda et al. 2010)? In any case to reach the necessary accuracy for constraining dark energy, one has to take into account some corrections which may depend on various parameters (Aubourg et al. 2008). So, any additional constraints on metallicity production by various types of heavy elements sources may occur to be very useful.

       It is also important to notice that, in the present paper, we use new much more precise determinations of oxygen and iron abundances in Cepheids.

                     
       \section {Basic ideas and equations}
       
       \subsection {{\it Equations for the formation of the fine structure of oxygen and iron radial distribution along the Galactic disc}}

       The chemical evolution of the Galactic disc is governed by the following equations:
       
       \begin{eqnarray}
       \dot{\mu}_O&=&\int\limits_{m_L}^{m_U}{(m-m_w)\,Z_O(t-\tau_m)\psi(t-\tau_m)\phi (m)\,dm}\nonumber\\
       &&\nonumber\\
       &&+E_O^{\rm II} + fZ_{Of} - Z_O\psi\nonumber\\ 
       &&\nonumber\\
       &&-\frac{1}{r}\frac{\partial {}}{\partial r}\left(r\mu_O u\right)
       +\frac {1}{r}\frac {\partial {}}{\partial r}
       \left(r\mu_g D\frac {\partial Z_O}{\partial r}\right),
       \end{eqnarray}

       \begin{eqnarray}
       \dot{\mu}_{Fe}&=&\int\limits_{m_L}^{m_U}{(m-m_w)\,Z_{Fe}(t-\tau_m)\psi(t-\tau_m)\phi (m)\,dm} \nonumber\\
       &&+E_{Fe}^{\rm Ia-P}+ E_{Fe}^{\rm Ia-T}+E_{Fe}^{\rm II} + fZ_{Fef} -
       Z_{Fe}\psi\nonumber\\ 
       && \nonumber\\
       &&-\frac{1}{r}\frac{\partial {}}{\partial r}\left(r\mu_{Fe} u\right)
       +\frac {1}{r}\frac {\partial {}}{\partial r}\left 
       (r\mu_g D\frac {\partial Z_{Fe}}{\partial r}\right), 
       \end{eqnarray}  
       where $\mu_O$ and $\mu_{Fe}$  are the surface mass densities for oxygen  and iron, respectively,
       $\mu_g$ is the gaseous density,
       $Z_i=\mu_i/\mu_g$ is the fraction of the $i$th element (oxygen or iron) in ISM,
       $Z_{Of}$ and $Z_{Fef}$ are the oxygen and iron abundances in the infall gas (for the both elements we adopt $Z_{if} = 0.02Z_{i\odot}$: our experiments with various abundances of the infall gas from $Z_{if}=0.02\,Z_{i\odot}$ to $0.1\,Z_{i\odot}$ show that the final abundances weakly depend on the exact value of $Z_{if}$ if it is less than $0.1\,Z_{i\odot}$, see also Lacey \& Fall 1985. Below we demonstrate the results for $Z_{if} = 0.02Z_{i\odot}$ which is slightly less than the mean content of heavy elements in halo stars, $\sim0.03\,Z_{i\odot}$, Prantzos 2008),

       \begin{equation}
       \psi=\nu\mu_g^{1.5}
       \end{equation}
       is the star formation rate (SFR), $\nu$ is a normalizing coefficient,
       $\phi(m)$ is Salpeter's initial mass function with the exponent of -~2.35 (stellar masses $m$ are in solar units),
       $E^{\rm II}_i$ are the rates of the $i$th element synthesis by SNe II,
       $E^{Ia-P}_{Fe}$ and $E^{Ia-T}_{Fe}$ are the rates of iron synthesis by prompt and tardy SNe Ia, respectively,
       \footnote {Acharova et al. (2010) showed that SNe Ia produce only about 2 per cent of oxygen (see also Matteucci 2004 and Tsujimoto et al. 1995). That is why we neglect by the contribution of SNe Ia to synthesis of oxygen.}
       $f$ is the infall rate of the intergalactic gas on to the Galactic disc 
       \begin{equation}
       f=A\exp(-\frac{r}{r_d}-\frac{t}{t_f}),
       \end{equation}
       $r_d = 3.5$ kpc is the radial scale, $t_f$ is a typical time-scale of gas fall on to the 
       Galactic disc, or in other words, the time-scale of the Galactic disc formation,
       $u$ is the microscopic radial gas velocity (radial inflow) within the Galactic disc,
       $t$ is time (in Gyr),
       $\tau_m$ is the life-time of a star of mass $m$ on the main sequence: $\log(\tau_m)=0.9-3.8\log(m)+\log^{2}(m)$,
       $m_L=0.1$, $m_U$ is the upper stellar mass (usually we adopt $m_U$ = 70, see below),
       $m_w$ is the mass of stellar remnants (white dwarfs, neutron stars, black holes: for $m \le 10$ the mass of a remnant is $m_w = 0.65m^{0.333}$; in the range  $10 < m < 30$ $m_w = 1.4$; if $30 \le m < m_U$ the remnant is a black hole with $m_w = 10$; finally for $m \ge m_U$ the stars are black holes right away from their birth and they are removed from the nucleosynthesis and returning the mass to ISM). The last terms on the right-hand sides of Eqs. (1,2) describe the turbulent diffusion of heavy elements with the diffusion coefficient $D$. To estimate the coefficient we model the turbulent ISM by a system of clouds and use the gas kinetic approach (for details see Mishurov et al. 2002; Acharova et al. 2010).

       The enrichment rates $E_i^{\rm II}$ of the ISM by the $i$th heavy element due to SNe II explosions are described by the same expressions: 
       \begin{equation}
       E_i^{\rm II}=\eta P_i^{\rm II} R^{\rm II},
       \end{equation}
       where $P_i^{\rm II}$ is the mean mass (in solar units) of ejected oxygen or iron per one SN II explosion, 
       \begin {equation}
       R^{\rm II}(r,t)=0.9975\int\limits_{8}^{m_U}{\psi(r,\,t-\tau_m)\phi (m)\,dm},
       \end {equation}
       is the rate of SNe\,II events. 
       
       The factor $\eta$ was introduced in order to take into account the influence of spiral arms. Following the idea, first proposed by Oort (1974) and used by Portinari \& Chiosi (1999) and Wyse \& Silk (1989), we write 
       
       \begin{equation}
       \eta=\beta |\Omega(r)-\Omega_P|\Theta, 
       \end{equation}
       where $\Omega(r)$ is the angular rotation velocity of the galactic disc, $\Omega_P$ is the rotation velocity of the wave pattern responsible for the spiral arms, $\Theta$ is a cutoff factor ($\Theta=1$ in the wave zone, i.e. between the inner and outer Lindblad resonances, and $\Theta=0$ beyond them), $\beta$ is a normalizing coefficient which we call as the constant for the rate of oxygen synthesis (for details see Mishurov et al. 2002; Acharova et al. 2005; 2010; 2011).

       Let us now turn to the synthesis of iron. In equation (2) the rates of enrichment of the Galaxy by iron due to SNe Ia-P and SNe Ia-T events are explicitly separated ($E_{Fe}^{\rm Ia-P}$ and $E_{Fe}^{\rm Ia-T}$, respectively). Being young objects, SNe Ia-P are believed to be concentrated in spiral arms. Hence, in addition to SNe II, they represent a complementary channel by means of which spiral arms influence the formation of multi-slope gradient of iron distribution in the disc. Therefore, by analogy with the representation for the enrichment rates due to SNe II events, $E_{Fe}^{\rm Ia-P}$ has to contain the factor $\eta$. So, the contribution of SNe Ia-P to the enrichment rate of the Galaxy by iron is governed by the following expressions: 
       \begin{equation}
       E_{Fe}^{\rm Ia-P}=\eta\gamma P^{\rm Ia}_{Fe} R^{\rm Ia-P},
       \end{equation}
       where $\gamma$ is a correction factor, $P^{\rm Ia}_{Fe}$ and $R^{\rm Ia-P}$ have the same sense as the corresponding quantities for the SNe II,
        
       \begin {equation}
       R^{\rm Ia-P}(r,t)= 0.00711\int\limits_{\tau_8}^{\tau_S}{\psi(r,\,t-\tau)D_P(\tau)d\tau},
       \end {equation}
       $D_P$ is the DTD function for prompt SNe Ia, $\tau_8$ is the life-time for a star of mass $m=8$.

       Unlike prompt SNe Ia, SNe Ia-T do not concentrate in spiral arms since their precursors are long-lived objects. That is why the $\eta$-like factor is absent in the expression for $E_{Fe}^{\rm Ia-T}$. Therefore, the contribution to iron enrichment of the ISM by tardy SNe Ia is described by the following formula: 
       \begin{equation}
       E_{Fe}^{Ia-T}=\zeta P^{\rm Ia}_{Fe}R^{\rm Ia-T}.
       \end{equation}
       Here unlike the above, $\zeta$ is a constant since this type of subpopulation of SNe Ia is not concentrated in spiral arms. So, they do not keep in their memory that they were born in spiral arms. The corresponding rate for SNe Ia-~T events is represented as follows:
       
       \begin {equation}
       R^{\rm Ia-T}(r,t)= 0.00711\int\limits_{\tau_S}^{t}{\psi(r,\,t-\tau)D_T(\tau)d\tau},
       \end {equation}
       where $D_T$ is the DTD function for SNe Ia-T.

       Let us specify, what we mean saying prompt SNe Ia. In Mannucci et al. (2006) and Matteucci et al. (2006) model the prompt and tardy subpopulations of SNe Ia are clearly separated since their DTD function is {\it bimodal}: the first group of objects has the delay time $\tau < \tau_S$, the second one corresponds to $\tau > \tau_S$. The critical time which serves as the boundary between the above subpopulations, $\tau_S$, happens to be $\sim 0.1$ Gyr [more exactly $\tau_S = 10^{7.93}$ yr $\approx 0.085$ Gyr, Matteucci et al. 2006; see equations (7,8) and figure 2 therein].

       On the other hand, the above critical time $\tau_S \sim 0.1$ Gyr can be considered as a boundary delay time which divides prompt SNe Ia from tardy ones in the case of {\it smooth} DTD function, proposed, e.g., by Maoz et al. (2010). Indeed, the typical time, necessary for a star to cross the interarm distance, is $\sim \pi /|\Omega - \Omega_P| \,>$~200 Myr (in the vicinity of the {\it corotation resonance}, where $\Omega \to \Omega_P$, the crossing time $\to \infty$). So, we may adopt that, if the age of SNe Ia progenitor (i.e., delay time $\tau$) is less than $\tau_S \sim 0.1$ Gyr the corresponding SNe Ia are concentrated in spiral arms. In other words, the objects belong to subpopulation SN Ia-P.
       
       The above division is very important: as it will be shown below, the multi-slope gradient of iron distribution along the Galactic disc may be explained by the influence of spiral arms only if a significant portion of SNe Ia is concentrated in spiral arms. Hence, we believe that SNe Ia-P have to be sufficiently young, no older than $\sim$100 Myr.

       Below we consider two types of approximating representations for the DTD function.
       
       1) {\it Bimodal DTD} function of Matteucci et al. (2006): $\log(D_P)=1.4-50{[\log(\tau)+1.3]}^2$) for $\tau \le \tau_S$ and $\log(D_T)=-0.8-0.9{[\log(\tau)+0.3]}^2$ for $\tau > \tau_S$, $\tau_S = 0.085$ Gyr.
       
       From the above representation it is seen that $D_P$ has a very sharp maximum at $\tau_{max} \approx 0.05$ Gyr due to the large parameter [$=50\ln(10) \approx 115$]. So, the main contribution to the integral for $R^{\rm Ia-P}$ brings the region of $\tau$ close to $\tau_{max}$. Hence, the integral may be estimated asymptotically by means of Laplace method.

       2) {\it Smooth DTD} function of Maoz et al. (2010). In this model we use a power-like DTD function ($DTD \propto \tau^{-1.2}$), proposed in the cited paper. However, we slightly modify it for small time, say, $\tau < 0.045$ Gyr assume $DTD$ to be proportional to $\exp \{-[(\tau - 0.04)/0.02]^2\}$ in order to avoid the step-like behavior of it at early times. Normalizing DTD to 1 within the time range from 20 Myr to 18 Gyr
       \footnote {The upper limit for $\tau$ is dictated by the least mass ($\sim$0.8 M$_{\odot}$) of a white dwarf companion in order the binary system results in SNe Ia outburst (see Greggio 2005; Matteucci et al. 2006). So, there is no any contradiction with the shorter age of the Universe.}
       and suppose that DTD is a continuous function, finally we have: $DTD = 0.135\tau^{-1.2}$ for $\tau \ge 0.045$ Gyr and $DTD = 5.940 \exp \{-[(\tau - 0.04)/0.02]^2\}$ for $\tau < 0.045$ Gyr. In this model, as SNe Ia-P we consider the stars for which the delay time $\tau < 100$ Myr. Otherwise we refer the stars to SNe Ia-T.

       As it was noticed in Introduction, Cepheids in our sample are very young: their ages usually do not exceed 100 Myr (see Table 1 in Appendix). So, they give the distribution of abundances in ISM almost at present epoch. And it is important, since the above chemical equations describe the evolution of the abundances of heavy elements in ISM. Therefore, by means of our chemical equations we have to compute the theoretical distributions of oxygen and iron for the present moment of time $t = T_D =10$ Gyr ($T_D$ is the age of the Galactic disc) and compare the theoretical distributions with the observed ones. To do that we have to transform our final $Z_i$ to metallicities: $[X_i/H]^{\rm th} = log(Z_i/Z_{i,\odot})$, where $Z_{i,\odot}$ is the abundance of oxygen or iron for the Sun. The corresponding values for $Z_{i,\odot}$ were adopted according to Asplund et al. (2009).

       The fundamental feature of the above equations for the chemical evolution of the Galactic disc is that they result in formation of the nontrivial radial distribution of the elements in the Galactic disc. Indeed, from the galactic density wave theory of Lin et al. (1969) it is known that the spiral wave pattern, responsible for spiral arms, rotates as a rigid body ($\Omega_P = const$) whereas the galactic matter rotates differentially (the rotation velocity of the Galactic disc $\Omega$ is a function of the Galactocentric distance $r$; to compute the rotation curve we use CO data of Clemens 1985, adjusted them for $r_{\odot} = 7.9$ kpc, see Acharova et al. 2010). The radius $r_c$, where both the velocities coincide [$\Omega(r_c) = \Omega_P$], is called the {\it corotation} radius. From the above expression for $\eta$ it is obvious that in the vicinity of the corotation radius the enrichment of ISM by SNe II and SNe Ia-P is depressed since here the difference $|\Omega-\Omega_P| \to 0$. The combined effect of the corotation resonance and turbulent diffusion results in formation of the radial distribution of heavy elements with the slope which varies along the galactocentric radius. 
       
       For completeness we also include in our theory the radial gas inflow within the Galactic disc [see the divergent terms in the last lines of equations (1,2)]. For the radial velocity $u(r)$ we adopt the same model representations as in the paper by Acharova et al. (2011).
       
       Equations (1,2) are the ones in partial derivatives. So, besides the initial conditions (at $t=0$ the initial values $\mu_i = \mu_g = 0$) we adopt the so-called natural conditions of the finiteness of the solutions at the Galactic center and at the Galactic disc end, $r_G$ (for models with radial gas inflow we locate the Galactic end at $r_G = 35$ kpc, in the case $u = 0$ the value $r_G = 25$ kpc is adopted).
       
       Strictly speaking the full system of equations for the Galactic chemical evolution includes also the equations for the disc formation, which describe the exchange by mass among the intergalactic matter, gaseous and stellar components. However, we do not write them since they and their solutions are the same as in Acharova et al. (2011, see figure 1 therein). We only notice that according to the last paper, the short time-scale of the Galactic disc formation, $t_f \sim 2$ Gyr, fits the best both to the observations of low present rate of gas infall on to the Galactic disc, $\sim$~0.1-0.2 M$_{\odot}$yr$^{-1}$, and the star formation rate which is expected to be of the order of magnitude higher (Sancisi \& Fraternali 2008, Bregman 2009, Robitaille \& Whitney 2010). So, for the Galactic disc formation we adopt the results of Acharova and collaborators [the values of constants $A$ and $\nu$ for various models of inflow, i.e. $u(r)$, see in Table 1 of their paper].

       \subsection {{\it Statistical method}}

       The above system of equations for chemical evolution of the Galaxy has 4 free parameters: $\beta$, $\Omega_P$, $\gamma$ and $\zeta$. To derive them we try to fit the theory to observations minimizing the merit function (or discrepancy) $\Delta$
       
       \begin{equation}
       \Delta^2 = \frac{1}{n-p}\sum_{i=1}^n\{(\langle[X/H]^{ \rm ob}\rangle_i - [X/H]^{ \rm th}_i)w_i\}^2
       \end{equation}
       over the above free parameters. Here 
       $[X/H] = log(N_X/N_H)_s - log(N_X/N_H)_{\odot}$, $N_X$ and $N_H$ are the numbers of atoms of the element $X$ and that of hydrogen in the object, respectively, the first term on the right-hand side of the last relation refers to a star, 
       the second one - to the Sun, the superscript `ob' corresponds to the observational and `th' to theoretical data, the symbol $\langle ... \rangle$ means that we apply our theory to a group of stars which fall into a bin centered at the $i$th Galactocentric radius $r$, $w_i$ is the weight, $n$ is the number of bins, $p$ is the number of the sought for free parameters, the summation is taken over all $i$th bins of the Galactocentric radius where the abundances of the elements were measured. 
       
       To estimate the errors of the sought for parameters we compute the confidence (at the level 95 \%) contour
       
       \begin{equation}
       \Delta_c^2 = \Delta_m^2 [1 + \frac{p}{n-p}F(p,n-p,0.95)],
       \end {equation}
       where $\Delta_m$ is the minimal value for the discrepancy, $F$ is Fisher's $F$ statistics (see Draper \& Smith 1981).

       Below we show that the process of the statistical treatment of observational data may be divided into two steps. Indeed, as it was noticed in Sec. 2.1 we can neglect by the contribution of SNe Ia to oxygen synthesis. Hence, the values $\Omega_P$ and $\beta$ can be derived independently of other two target parameters $\gamma$ and $\zeta$ since the last 2 quantities do not enter the corresponding equations describing oxygen production.

       So, at {\it Step 1} we analyze the oxygen distribution to evaluate $\Omega_P$ and $\beta$. For this, we solve equations (1,3-7) for a set of $\Omega_P$ and $\beta$. After that we construct the surface $\Delta$ as a function of $\Omega_P$ and $\beta$ ($p = 2$), find out the minimum of $\Delta$ which determines the best values for the above parameters and compute the corresponding confidence contour for them.

       At {\it Step 2} the radial distribution of iron is analyzed. Now we seek for the last 2 free parameters, $\gamma$ and $\zeta$. The idea for evaluation them is similar to the method used at the previous step, that is, we solve equations (2-11), describing iron synthesis, for a set of $\gamma$ and $\zeta$, consider $\Omega_P$ and $\beta$ as have been derived at previous step, than compute the discrepancy between the theoretical and observed distributions of iron as a function of $\gamma$ and $\zeta$, again construct the surface of $\Delta$, but now as a function of $\gamma$ and $\zeta$, look for its minimum which gives the best values for $\gamma$ and $\zeta$. 
       
       However, at this step we have to take into account that the equations, describing the synthesis of iron, contain the parameters $\Omega_P$ and $\beta$, obtained independently at the previous step. Therefore we have to take into account the influence of their errors on biases and errors in $\gamma$ and $\zeta$. For this, we propose a kind of numerical experiment (see Sec. 4).

       Above we discussed the methods of estimation the statistical errors for the free parameters. But there is a source of errors which have another nature, namely: the uncertainties in oxygen and iron yields. As starting values, in our computations, we adopt the masses of oxygen and iron, ejected per one SN event, from Tsujimoto et al. (1995): 
       $P^{\rm II}_{O} = 2.47$, $P^{\rm II}_{Fe} = 0.084$, and  $P^{\rm Ia}_{Fe} = 0.613$ (following Matteucci et al. 2006, for the both SNe Ia subpopulations here we use the same ejected masses of iron). On the other hand, in literature one can find other values for the ejected masses (see, e.g., Woosley \& Weaver 1995, Thielemann 1996 and others). Besides prompt and tardy SNe Ia may have different outputs of iron (Howell et al. 2009). How do the changes in the ejected masses influence the final amounts of the elements supplied by various Types of SNe to the Galactic disc? 
       
       To feel the answer this question, first, let us look at the structure of the rate for oxygen enrichment: from equation (5) it is seen that $P_O^{\rm II}$ enters the expression for $E^{\rm II}_{O}$ as a product $\beta P_{O}^{\rm II}$. Similarly, the enrichment rate by iron due to SNe II explosion is proportional to the product $\beta P_{Fe}^{\rm II}$. Hence, if we adopt another value for $P_{O}^{\rm II}$,  the constant $\beta$ will change, so that the product $\beta P_{O}^{\rm II}$ to be kept the same in order the final amount of oxygen to be unalterable. But this will influence the enrichment rate by iron due to SNe II even if $P_{Fe}^{\rm II}$ is retained unchanged.
       
       In turn, the enrichment rates for iron by SNe Ia are proportional to products $\beta \gamma P_{Fe}^{\rm Ia-P}$ for prompt and $\zeta P_{Fe}^{\rm Ia-T}$ for tardy objects [see equations (8,10)]. So, it is obvious, the variation in mass of oxygen (!), ejected during SNe II explosions, influences the output of iron due to SNe Ia-P, since the corresponding enrichment rates by iron for SNe II and SNe Ia-P have close functional representations along the Galactic radius. But the amount of iron supplied by SNe Ia-T does not change. In other words, in this case, we should have a redistribution of amount of iron among the 3 sources of it.

       Consider the second possible case: $P_O^{\rm II}$ is equal to the starting value but $P_{Fe}^{\rm II}$ is changed. Hence, the enrichment rate $E_{Fe}^{\rm Ia-P}$ has to be inversely changed in order to compensate the variation in the rate of iron enrichment due to SNe II, but again we do not expect any significant variations in the amount of iron synthesized by SNe Ia-T.
       
       At last, in the third case, let us consider the result if $P_{Fe}^{\rm Ia}$ for SNe Ia-P and SNe Ia-T are different, but other ejected masses are equal to the starting values. It is easy to see that the final amounts of iron supplied by all Types of SNe will not change at all. Only constants $\gamma$ and $\zeta$ will alter in order the products $\gamma P_{Fe}^{\rm Ia-P}$ and $\zeta P_{Fe}^{\rm Ia-T}$ do not change relative to the starting case.

       In Sec. 4 we illustrate the discussion of this problem by some results of our numerical experiments.

       After evaluation of the free parameters we compute the amount of iron synthesized by each type of its sources.

                     
       \section {Observational data}

       In the present paper, we use the most extensive spectroscopic (only) data
       on oxygen and iron abundances derived for classical 283 Cepheids (872 spectra in total). A part of the data were previously published (\cite{lu11} and references therein). For completeness we give the data in Table 1 (see Appendix).
       Below we describe briefly our methods and analysis of spectra.
       
       
       \subsection{Spectral material}
       
       The spectra of additional Cepheids were obtained using the facilities of the 1.93\,m telescope at the  Haute-Provence Observatoire (France) equipped  with 
       \'echelle-spectrograph ELODIE. In the region of wavelengths 4400--6800 \AA\, the resolving power was R=42\,000, the signal-to-noise ratio, S/N, being about 80--130. The initial processing of the spectra (image extraction, cosmic particles removal, flatfielding, etc.) was carried out following to Katz et al.
       (1998). Also we use \'echelle-spectrograph SOPHIE at this telescope, the spectra stretch from 3870 to 6940 \AA \,in 39 orders with resolution R=75\,000.
       
       Some spectra of Cepheids were obtained with the fiber \'echelle-spectrograph
       HERMES mounted on the 1.2\,m Belgian telescope on La Palma. A high-resolution configuration with R= 85\,000 and wavelength coverage 3800--9000 \AA \,is used. The spectra were reduced using a Python-based pipe-line, following a procedure of the order extraction, wavelength calibration using Thr-Ne-Ar arcs, division
       by the flat field, cosmic-ray clipping, and the order merging. For more details on the spectrograph and the pipe-line see Raskin et al. (2011).
       
       We also made use of spectra obtained with the Ultraviolet-Visual Echelle
       Spectrograph (UVES) instrument at the Very Large Telescope (VLT) Unit 2
       Kueyen (Bagnulo et al. 2003). All supergiants were observed in two instrumental modes, $Dichroic \,1$ and $Dichroic \,2$, in order to provide almost complete coverage of the wavelength interval 3000-10\,000 \AA. The spectral resolution is about 80\,000, and for most of the spectra the typical S/N ratio is 150--200.
       
       Further processing of the spectra (continuum level location, measurement of the equivalent widths, etc.) was performed using the software package DECH20 (Galazutdinov 1992). The equivalent widths were measured using the Gaussian fitting.
       
       \subsection{{\it Atmospheric parameters}}
       
       Effective temperatures for our program stars were established from the processed spectra using the method developed by \citet{ko07} that is based upon $T_{\rm eff}$--line depth relations. The technique can establish $T_{\rm eff}$ with exceptional precision. It relies upon the ratio of the central depths of two lines that have very different functional dependences on $T_{\rm eff}$, and uses tens of pairs of lines for each spectrum. The method is independent of interstellar reddening, and only marginally dependent on the individual characteristics of stars, such as rotation, microturbulence, metallicity, etc.
       
       The microturbulent velocities, $V_{\rm t}$, and surface gravities,
       $\log g$, were derived using a modification of the standard analysis proposed by \citet{ka99}. As described there, the microturbulence is determined from the Fe~II lines rather than the Fe~I lines, as in classical abundance analyses. The surface gravity is established by forcing equality between the total iron abundance obtained from both Fe~I and Fe~II lines. Typically with this method the iron abundance determined from Fe~I lines shows a strong dependence on equivalent width (NLTE effects), so we take as the proper iron abundance the extrapolated total iron abundance at zero equivalent width.
       
       Kurucz's WIDTH9 code was used with an atmospheric model for each star interpolated from a grid of models calculated with a microturbulent velocity of 4 km s$^{-1}$ Kurucz (1992). At some phases Cepheids can have microturbulent velocities deviating significantly from that value; however, our previous test calculations suggest that changes in the model microturbulence over a range of
       several km s$^{-1}$ has an insignificant impact on the resulting element abundances. The oscillator strengths used in this and all preceding Cepheid analyses of this series are based on an inverted solar analysis.

       \subsection {{\it Distances, masses and ages of the Cepheids}}
       
        The heliocentric distance, $d$, of a Cepheid is estimated in a usual way:
        
       $$d = 10^{-0.2 (M_{\rm v} - <V> -5 + A_{\rm v})},$$
       where $M_{\rm v}$ is the absolute magnitude, $<V>$ is the mean visual magnitude, $A_{\rm v}$ is the line of sight extinction,
       $A_{\rm v} = 3.23 E(B-V)$ (pulsate periods, mean visual magnitudes, colors and $E(B-V)$ values are taken from \citet{fernie95}; $M_{\rm V} - P$ relation from \citet{fouque07} and \citet{ko08}). To transform $d$ to the galactocentric distance, $r$, of the Cepheid we use the Galactocentric solar distance $r_0 = 7.9$ kpc.
       
       The masses and ages for our Cepheids are derived using $Period - Mass$ relation from \citet{turner96} and $Period - Age$ relation from \citet{Bono05}. Our estimates show that the ages of the most portions of Cepheids from our sample are less than 100 Myr. Only several stars are older, but in any case their ages do not exceed 130 Myr. Hence they did not undergo significant radial scattering. So, all Cepheids demonstrate the distribution of the elements in the ISM, almost at the Galactocentric radius, where we observe them at the present moment of time.
       
       The table with the derived abundances and other parameters for all Cepheids is given in Appendix.

       \subsection {Radial distributions of oxygen and iron along the Galactic disc}
       
       For modeling, we divide the galactocentric radius in bins of some width and average the abundances within the bins over the stars which have fallen to the bin. As in our previous papers, in Fig.1 we show the radial distributions of the mean abundances for oxygen, $\langle [O/H]^{ob}\rangle$, and iron, $\langle [Fe/H]^{ob}\rangle$, and their relation along the Galactic disc at step of 0.25 kpc, the bin width being equal to 0.5 kpc. Bars in the figure describe the scatter of the above mean abundances within the bin. In our statistic analysis, we adopt the weight $w_i$ [see equation (1)] to be inversely equal to the length of the bar in the $i$th bin.

       \begin{figure}
              \includegraphics {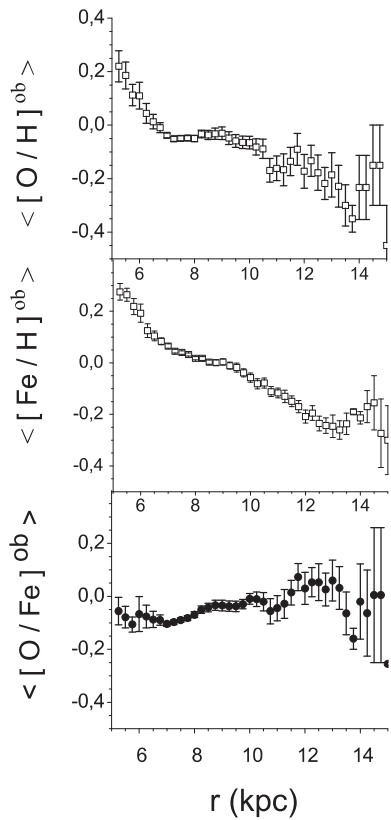}
              \caption{Radial distributions of oxygen, iron and their relation along the Galactic disc, averaged over bins of 0.5 kpc width. Bars correspond to the scatter of the mean abundance (see text for details).}
              \label{f1}
              \end{figure}

       Let us discuss some features of the distributions in figure 1. First of all, notice that the scatter of the mean abundances happens to be much less than the one which was computed on the basis of previous observational data of 
       Andrievsky et al. (2002 a,b,c) and Luck et al. (2003, 2006; see figure 1 in Acharova et al. 2010). Such decrease in the scatter is obviously a result of improvement of the abundances determinations and increase in number of objects. However, at large Galactocentric distances the scatter happens to be much greater than in the inner region.
       
       Further, the radial distribution of oxygen demonstrates sufficiently sharp inflection of the slope in the distribution at $r \sim 7$ kpc. But for iron there is no such sharp bend in the gradient at the same Galactocentric distance. Nevertheless, it is obvious that the radial distribution of iron cannot be satisfactorily described by a trivial linear function.

       At last, the distribution of iron is rather smooth up to $r \sim 13$ kpc.  New data do not show any visible gaps or jumps for smaller radii. The increase followed by the decrease in its content between 13 and 15 kpc takes place at approximately the same distance that of for oxygen although it is not so prominent and the both radial patterns differ in details. For instance, there is no noticeable variation in iron distribution at $r \sim 10.5$ kpc where in oxygen distribution we see a rather sharp step-like decrease. In our opinion, the above peculiarities may be associated with some local effects, say, with a sudden fall of a pristine gas on to the Galactic disc at $r \sim 10-11$ kpc or due to the Magellanic Stream at $r \sim 10-15$ kpc. However, we will not try to explain them: in spite of our model is difficult from the mathematical point of view such local effects cannot be simply incorporated in our theory. That is why for the statistical analysis we restrict ourselves by the region $r \le 10$ kpc. So, the number $n$ in equations (1,2) is $n = 20$. Since at the both steps (oxygen and iron analysis) $p =2$, Fisher's statistics $F(2,18,0.95) \approx 3.55$ (Draper \& Smith 1981).

       
       \section {Results and discussion}

       \subsection {\it{Step 1: Oxygen}}
       
       We performed calculations for the same models of the radial gas inflow [i.e., for the dependences of $u(r)$ of Acharova et al. (2011)] and the ejected mass of oxygen, $P_O^{\rm II}$ per one SNE II event of Tsujimoto et al. (1995). Unlike our previous paper, new observational data unambiguously lead to the least value for the discrepancy $\Delta$ (for oxygen we denote it as $\Delta^{\rm O}_m$) which corresponds to the model with no radial inflow (in notations of Acharova et al. 2011 it is the model `M20' with $u = 0$): $\Delta^{\rm O}_m = 0.641$
       \footnote {The increase of $\Delta^{\rm O}_m$ relative to the corresponding value of Acharova et al. (2011) is associated with that the weight $w_i > 1$.}
       which corresponds to the best values of $\Omega_P = 33.4$ km s$^{-1}$ kpc$^{-1}$ and $\beta = 0.0126$ Gyr. Comparing the above parameters with the ones from the last paper we see that the rotation velocity for the spiral density waves occurs to be the same (correspondingly, the corotation radius $r_c \approx 7$ kpc) whereas the coefficient $\beta$ has decreased by about 20 per cent. Besides, the confidence contour has changed distinctly (see figure 2): now the axes of the ellipse are not parallel to the axes of $\Omega_P$ and $\beta$. To indicate the confidence borders for the target parameters we adopt the following lower and upper values for them: $\Omega_P = 32.9\, -\, 34.2$ km s$^{-1}$ kpc$^{-1}$ (correspondingly $r_c = 7.1\, -\, 6.8$ kpc);  $\beta = 0.0129\, -\,  0.0122$ Gyr (in figure 2 they are labeled by filled circles marked as `A' and `B'). For simplicity we adopt the symmetrical errors in $\beta$, so finally $\beta = 0.0126 \pm 0.0004$ Gyr.

       \begin{figure}
              \includegraphics {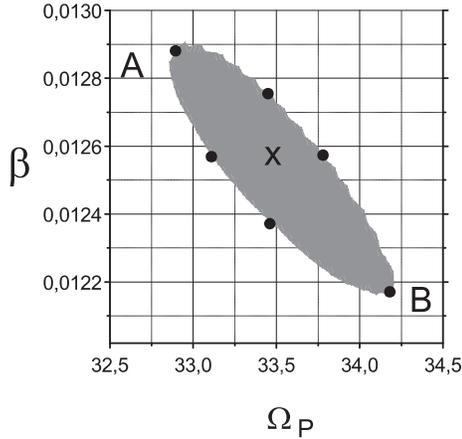}
              \caption{The confidence contour for $\Omega_P$ and $\beta$. The cross corresponds to the best values of $\Omega_P$ and $\beta$.}
              \label{f2}
              \end{figure}

       In figure 3 we show the theoretical radial distributions of oxygen computed for the best above parameters and for their values, corresponding to the extreme points of the confidence contour from figure 2 (`A' and `B'), superimposed on the observational distribution. Within the radius range $5 \le r \le 10$ kpc the coincidence of the theory with observations is excellent. Notice the very good agreement of the theory with observations both at $r \sim 7$ kpc where there is the bend in the gradient slope and in the range of the flat (a plateau-like) oxygen distribution.

              \begin{figure}
              \includegraphics {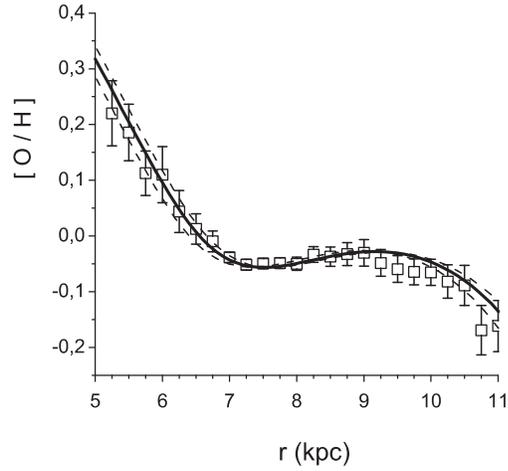}
              \caption{The comparison of the theoretical radial distribution of oxygen with the observations. {\it Solid} line is for the best values of $\Omega_P,\, \beta$ which correspond to the cross in figure 1; {\it dashed} lines correspond to the parameters labeled by points `A' and `B' in the previous figure.}
              \label{f3}
              \end{figure}

       Our computer experiments confirm the statement, made in Sec. 2.1, that the radial distribution of oxygen, its full synthesized mass and $\Omega_P$ do not change if we adopt another value for $P_{O}^{\rm II}$. Only $\beta$ alters but so as the product $\beta P_{O}^{\rm II}$ has to be kept the same.

       \subsection {\it{Step 2: Iron}}

       Oxygen is mainly synthesized during SNe II events. So, it is the most pure indicator of spiral arms influence on heavy elements synthesis in the Galactic disc. Besides, since we can neglect by the contribution of SNe Ia to its abundance, of the discussed two elements the kinetics of oxygen synthesis is simpler. 
       
       Unlike oxygen, iron is synthesized by SNe II, SNe Ia-P and SNe Ia-T. To estimate the contributions of each type of the above sources to the production of iron is more difficult problem than the one for oxygen. Indeed, to solve the posed task we have to derive the constants for the rates of iron synthesis by means of fitting our theory to the observed fine structure of the radial distribution of iron in the Galactic disc.

       Now we set out, in short, our method for evaluation of the free parameters $\gamma$ and $\zeta$. For this, first of all, notice that the enrichment rate of ISM by iron due to SN II explosions is described by the same relation (5) with the only substitution: $P_{i}^{\rm II} \rightarrow P_{Fe}^{\rm II}$. Since the constant $\beta$ was derived at {\it Step 1} the contribution of SN II to iron synthesis is determined entirely. Hence, we only need to derive the constants $\gamma$ and $\zeta$ [see equations (8,10)] by means of fitting the theory to the observed radial distribution of iron. In a general way, the procedure of evaluation them is similar to the one for derivation of $\Omega_P$ and $\beta$, namely, for the fixed values of $\Omega_P$ and $\beta$ we solve numerically the equations of iron synthesis in the Galactic disc, varying the sought for parameters $\gamma$ and $\zeta$. Then, using the theoretical and observational data for the radial distribution of iron, we again compute the net of the merit function $\Delta^{\rm Fe}$ (the superscript `Fe' means that the discrepancy refers to iron) as a function of $\gamma$ and $\zeta$, find out its minimum  over $\gamma$ and $\zeta$ ($min \, \Delta^{\rm Fe} = \Delta^{\rm Fe}_m$) and derive the confidence contour for them. To control the process, we construct the surface $\Delta^{\rm Fe} (\gamma, \zeta)$, an example of which is given in figure 4 for the best values of $\Omega_P$ and $\beta$. In figure 5 we demonstrate the corresponding confidence contour for $\gamma$ and $\zeta$ computed for the best values of $\Omega_P$ and $\beta$. However, at this step we have to take into account that the quantities $\Omega_P$ and $\beta$, derived at the previous step, have errors which may result in biases of the best values of $\gamma$ and $\zeta$ and their errors. To solve this problem we made a numerical experiment, repeating the above procedure for various values of $\Omega_P$ and $\beta$ from their confidence contour and averaging $\gamma$ and $\zeta$ and their errors. 
       
       Our computations show that $\gamma$ and $\zeta$ and the disposition of the confidence ellipse for them change slightly if we use $\Omega_P$ and $\beta$ from their confidence region. Therefore, we test only the 7 pairs of values $\Omega_P$ and $\beta$ which are labeled by filled circles and by the cross in figure 2. 
       
       \begin{figure}
       \includegraphics {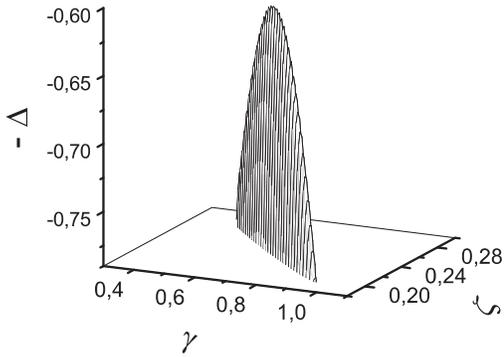}
       \caption {An example of the surface $\Delta^{\rm Fe} (\gamma, \zeta)$ for the best values of $\Omega_P$ and $\beta$. For better visualization we draw the surface `bottom-up'.}
       \label{f4}
       \end{figure}

       \begin{figure}
       \includegraphics {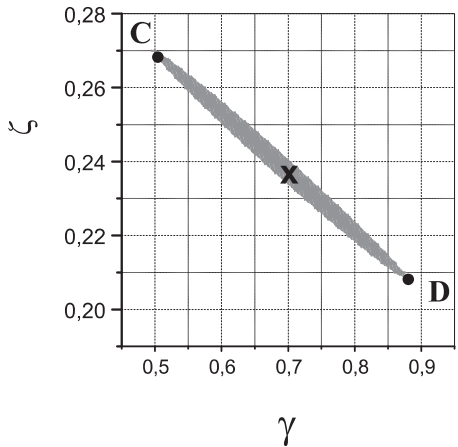}
       \caption {An example of the confidence contour for $\gamma$ and $\zeta$ computed for the best values of $\Omega_P$ and $\beta$. The cross corresponds to $\Delta^{\rm Fe}_m$ for the above values of $\Omega_P$ and $\beta$.}
       \label{f5}
       \end{figure}

       As a result, we derive 7 pairs of values for $\gamma$ and $\zeta$ corresponding to a particular minimum $\Delta^{\rm Fe}_m$ (denote them as $\gamma_m$ and $\zeta_m$) and to the largest deviations of $\gamma$ and $\zeta$ corresponding to the points `C' and `D' in the last figure [denote them as ($\gamma_C$, $\zeta_C$) and ($\gamma_D$, $\zeta_D$)]. Averaging ($\gamma_m$,\,$\zeta_m$) over the computed 7 results we find out the best values for the sought for parameters [denote them as ($\langle \gamma_m \rangle$, $\langle \zeta_m \rangle $]. By analogy we compute the mean extremal confident values for them, correspondingly, the pairs ($\langle \gamma_C \rangle$, $\langle \zeta_C \rangle$) and ($\langle \gamma_D \rangle$, $\langle \zeta_D \rangle$)). 
       
       In figure 6 is shown the comparison of the theoretical radial distribution of iron with the observations computed for the ejected mass of oxygen and iron from Tsujimoto et al. (1995) and supposing that SNe Ia-P and SNe Ia-T eject the same mass of iron per one event (i.e., $P_{Fe}^{Ia-P} = P_{Fe}^{Ia-T}$, see Sec. 2.2). Notice that in this figure we demonstrate the theoretical curve up to $r = 13$ kpc despite for the statistic analysis was treated only the region within 10 kpc. Nevertheless the agreement of our theory with the observations happens to be very good even in the extended region of Galactocentric radius.

       \begin{figure}
       \includegraphics {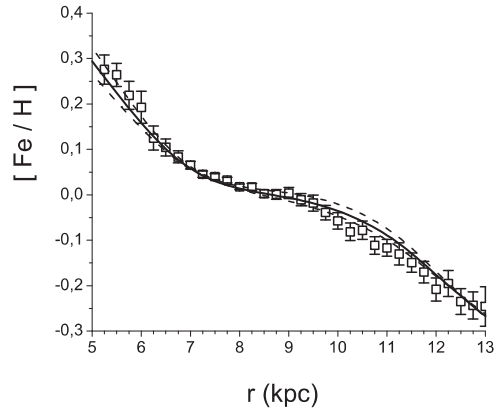}
       \caption {Theoretical radial distribution of iron superimposed on the observational data. 
       {\it Solid line} corresponds to the theoretical distribution computed for ($\langle \gamma_m \rangle$, $\langle \zeta_m \rangle$); 
       {\it dashed lines} correspond to ($\langle \gamma_{C} \rangle$, $\langle \zeta_{C} \rangle$) and ($\langle \gamma_{D} \rangle$, $\langle \zeta_{D} \rangle$).}
       \label{f6}
       \end{figure}

       In figure 7 is shown the theoretical relation for $[O/Fe]$ vs Galactocentric radius superimposed on the observations. As it was expected, the coincidence of the theory with observations again is good. This result independently demonstrates that the corotation resonance is located at the minimum of the radial distribution of the relation of oxygen to iron.

       \begin{figure}
       \includegraphics {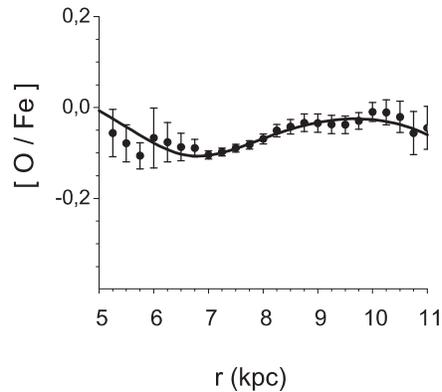}
       \caption {Comparison of the theoretical radial distribution of the relation $[O / Fe]$ with observations for the best values of $\beta$, $\gamma$, $\zeta$ and $\Omega_P$. Notice that the minimum of the relation of oxygen to iron at $r \sim 7$ kpc fits to the location of the corotation resonance.}
       \label{f7}
       \end{figure}

       Let us now discuss the effects of using DTD function of Maoz et al. (2010) which, unlike the bimodal function of Mannucci et al. (2006) and Matteucci et al. (2006), is smooth and peaked at early times (the corresponding approximation for the smooth DTD function see in Sec. 2.2). In figure 8 is shown the radial distribution of iron computed for the above smooth DTD function and the same ejected masses adopted from Tsujimoto et al. (1995). It is seen that the distribution differs slightly from the one derived for the bimodal DTD of Mannucci et al. (2006) and Matteucci et al. (2006).
       
       \begin{figure}
       \includegraphics {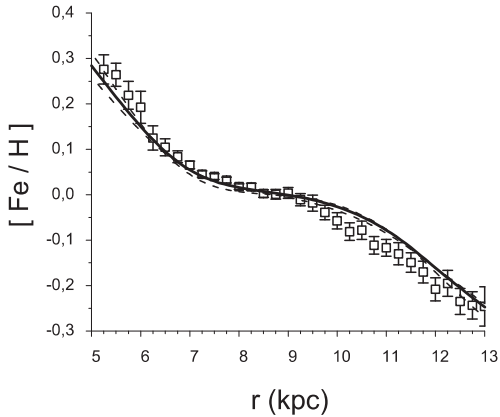}
       \caption {The same as in figure 6 but for the smooth DTD function of Maoz et al. (2010).}
       \label{f8}
       \end{figure}

       \subsection {Amount of iron synthesized by various sources}

       Now we can answer the question: how much iron is synthesized by each Type of SNe during the life-time of our Galaxy?

       Below we see that the mass of iron, synthesized by each Type of SNe for the period of life of the Galactic disc, differs from the corresponding masses which are kept in the present ISM. 
       \footnote {We neglect by the mass of iron which have fallen on to the Galactic disc with the infall gas during the life of the Galaxy since this mass happens to be $\sim~2.2\cdot 10^5$ M$_{\odot}$ and is much less than the mass of iron produced by SNe.}
       Hence, saying the amount of iron supplied to the Galaxy by any Type of SNe we mean the mass which was synthesized by the corresponding SNe Type for the age of the Galactic disc. To compute these quantities (denote them as $M_{Fe}^{\rm II}$, $M_{Fe}^{\rm Ia-P}$ and $M_{Fe}^{\rm Ia-T}$) we simply integrated the corresponding enrichment rates [see eqs. (5,8,10)] over the surface of the Galactic disc and time. 
       
       However, the procedure for evaluation of the above masses of iron which occur in the present ISM differs from the one, described before. Indeed, equation (2) governs the evolution of iron content in ISM. So, to find out $M_{Fe}^{\rm II}$, $M_{Fe}^{\rm Ia-P}$ and $M_{Fe}^{\rm Ia-T}$ in the present ISM, we have to solve the corresponding equations separately for each type of iron sources, using the constants of the rates of iron synthesis evaluated at steps of fitting our theory to observations, and then integrate the derived $\mu_{Fe}(r,T_D)$ over the Galactic disc.

       In all experiments, considered by us, the radial distributions of iron along the Galactic disc are very close to each other and the distribution shown in figures 6 and 7. That is why we do not demonstrate the distributions derived for other input parameters and restrict our discussion by numerical values given in Table 2. Let us consider them in some details.

       
       \setcounter{table}{1}
       \begin{table*}
       \begin{minipage}{14cm}
       \caption{Constants $\beta$ (in Gyr), $\gamma$ and $\zeta$ (dimensionless) for the rates of synthesis of iron and masses, $M_{Fe}$ (in 10$^7$ M$_{\odot}$), synthesized by various Types of SNe during the age of the Galactic disc (columns denote by `{\bf Galaxy}') and confined in the present interstellar medium (columns denote by `{\bf ISM}'). Their random errors are given in parenthesis in the second row of {\it Case} 1). The ratio of the masses of iron supplied by various Types of SNe is given in per cent for each {\it Case}}
       \label{tab2}
       \begin{tabular}{cccccccccccc}
       \hline
       \multicolumn{4}{r}{}& \multicolumn{1}{c}{\vline} & \multicolumn{3}{c}{{\bf Galaxy}} & \multicolumn{1}{c}{\vline} &
       
       \multicolumn{3}{c}{\bf ISM} \\
       \hline

         &$\beta$& $\gamma$& $\zeta$ & \vline & $M_{\rm Fe}^{II}$& $M_{\rm Fe}^{Ia-P}$& $M_{\rm Fe}^{Ia-T}$ & \vline & $M_{\rm Fe}^{II}$ & $M_{\rm Fe}^{Ia-P}$ & $M_{\rm Fe}^{Ia-T}$\\
       \hline
       Case 1: &  0.0126& 0.67 &  0.24 & \vline & 0.75 & 1.80 &  1.43 & \vline & 0.18 & 0.43 &  0.59 \\
         &$(\pm$0.0004)&  ($\pm$0.2)& ($\pm$0.03)& \vline & ($\pm$0.02) & ($\pm$0.60) & ($\pm$0.18) & \vline & ($\pm$0.01) & ($\pm$0.14) & ($\pm$0.08)\\
       
       \multicolumn{4}{c}{} & \vline& 19~\% & 45~\% & 36~\% & \multicolumn{1}{c}{\vline} & 15~\% & 36~\% &  49~\% \\
         
       \hline   
       
       Case 2: & 0.0174 & 0.44 & 0.24 & \multicolumn{1}{c}{\vline} & 1.03 & 1.63 & 1.43 & \multicolumn{1}{c}{\vline} & 0.25 & 0.39 & 0.59 \\
        &  &  &  & \multicolumn{1}{c}{\vline} & 25~\% & 40~\% & 35~\% & \multicolumn{1}{c}{\vline} & 20~\% & 31~\% & 49~\% \\
       
       \hline
       Case 3: & 0.0084 & 1.2 & 0.24 & \multicolumn{1}{c}{\vline} & 0.50 & 2.16 & 1.43 & \multicolumn{1}{c}{\vline} & 0.12 & 0.51 & 0.59 \\
        & & &  & \multicolumn{1}{c}{\vline} & 12~\% & 53~\% & 35~\% & \multicolumn{1}{c}{\vline} & 9~\% & 42~\% & 49~\% \\
       
       \hline               
       
       Case 4: & 0.0126 & 0.49 & 0.24 & \multicolumn{1}{c}{\vline} & 1.25 & 1.32 & 1.43 & \multicolumn{1}{c}{\vline} & 0.30 & 0.31 & 0.59 \\
        &  &  & & \multicolumn{1}{c}{\vline} & 32~\% & 33~\%& 35~\% & \multicolumn{1}{c}{\vline} & 25~\% & 26~\%& 49~\% \\
       
       \hline
       
       Case 5: & 0.0126 & 0.54 & 0.32 & \multicolumn{1}{c}{\vline} & 0.75 & 1.81 & 1.43 & \multicolumn{1}{c}{\vline} & 0.18 & 0.43 & 0.59 \\
       &  &  &  & \multicolumn{1}{c}{\vline} & 18~\% & 44~\% & 38~\% & \multicolumn{1}{c}{\vline} & 15~\% & 36~\% & 49~\% \\
       
       \hline
       Case 6: & 0.0126 & 1.64 & 0.15 & \multicolumn{1}{c}{\vline} &  0.75 & 1.81 & 1.60 & \multicolumn{1}{c}{\vline} & 0.18 & 0.43 & 0.59 \\
        & &  &  & \multicolumn{1}{c}{\vline} & 18~\% & 44~\% & 38~\% & \multicolumn{1}{c}{\vline} & 15~\% & 36~\% & 49~\% \\

       \hline
       \hline
       \multicolumn {12}{l}{Case 1. Bimodal DTD of Matteucci et al. (2006); ejected masses are: $P_O^{\rm II}$ = 2.47; $P_{Fe}^{\rm II}$ = 0.084; }\\ \\
       \multicolumn {6}{l}{} & \multicolumn {6}{l}{$P_{Fe}^{\rm Ia-P}$ =$P_{Fe}^{\rm Ia-T}$ = 0.613 - Tsujimoto et al. (1995).}  \\ \\

       \multicolumn {12}{l}{Case 2. The same as in Case 1, but $P_O^{II}$ = 1.8, see text.} \\ \\

       \multicolumn {12}{l}{Case 3. The same as in Case 1, but $P_O^{II}$ = 3.7, see text.}\\ \\

       \multicolumn {12}{l}{Case 4. The same as in Case 1, but $P_{Fe}^{II}$ = 0.14 -  Thielemann et al. (1996)} \\ \\
       
       \multicolumn {12}{l}{Case 5. The same as in Case 1, but $P_{Fe}^{Ia-P}$ = 0.762; $P_{Fe}^{Ia-T}$ = 0.456, see text}\\ \\

       \multicolumn {12}{l}{Case 6. Smooth DTD function of Maoz et al. (2010); ejected masses are the same as in Case 1.} \\

       \end{tabular}
       \end{minipage}
       \\
       
       \end{table*}

       \subsubsection{ {\it Bimodal DTD function of Matteucci et al. (2006)}}

       In {\it Cases} 1 - 5 we examine the bimodal DTD function of Matteucci et al. (2006) varying the ejected masses of oxygen or iron during SNe events and analyzing the results of such changes. The input parameters in {\it Case 1} are considered as starting ones. For them we adopt the ejected masses per one SNe event from Tsujimoto et al. (1995) and outputs of iron from SNe Ia-P and SNe Ia-T are assumed to be the same (Matteucci et al. 2006). In the following 4 {\it Cases} we estimate the effects of variations of $P_{O}^{\rm II}$ or $P_{Fe}^{\rm II}$ and the supposition that $P_{Fe}^{\rm Ia-P} \ne P_{Fe}^{\rm Ia-T}$. Thus, in {\it Case} 2 we make an experiment with $P_O^{II}$ = 1.8: this value was proposed by Tsujimoto et al. (1995) for the upper stellar mass $m_U$ = 50 M$_{\odot}$ (in other {\it Cases} we use $m_U$ = 70 M$_{\odot}$). To illustrate the effect of increase of the mass of oxygen ejected by one SNe II on iron output, in {\it Case} 3 we compute the amounts of iron for $P_O^{II}$ = 3.7. This value is about 1.5 times greater than the starting one (notice that the ejected masses, derived by Woosley \& Weaver 1995, are systematically greater than the corresponding value of Tsujimoto et al. 1995 just about 1.5 times). At last, in {\it Case} 5 we compute the final masses of iron if we adopt that $P_{Fe}^{\rm Ia-P} \ne P_{Fe}^{\rm Ia-T}$. However, since the completed theory for the two subpopulation of SNe Ia is not built, we use for illustration the largest and the least values from Nomoto et al. (1997).

       In the second row of {\it Case} 1, in parenthesis, are shown the random errors of $\beta$, $\gamma$ and $\zeta$ evaluated by means of our statistical method and the errors for the masses of iron following from the above random errors in the constants for the rates of enrichment of the Galaxy by iron and oxygen. In other {\it Cases} the errors happen to be the same order of magnitude and we do not demonstrate them. Moreover, as it is seen from Table 2, sometimes the variations in the synthesized masses of iron due to uncertainties in the ejected masses, especially in $P_O^{\rm II}$ and $P_{Fe}^{\rm II}$, lead to greater variations in $M_{Fe}^{\rm II}$ and $M_{Fe}^{\rm Ia-P}$, although for long-lived SNe Ia progenitors the final value of $M_{Fe}^{\rm Ia-T}$ is sufficiently stable.
       
       The portions of iron masses supplied by various SNe to the Galactic disc and ISM are shown in per cent.

       It is interesting to notice that in all {\it Cases} the portion of mass of iron, synthesized by SNe Ia-T does not vary and happens to be equal to about 35 per cent. Correspondingly, the total portion of iron, produced by SNe II and SNe Ia-P, is $\sim$65 per cent. The only effect of the changes in ejected mass consists in redistribution of iron between SNe II and SNe Ia-P. This result confirms our suppositions made in Sec. 2.2.
       
       The same situation holds for the abundance of iron in the present ISM: about 49 per cent of it was supplied by SNe Ia-T, other 51 per cent were captured from SNe II and SNe Ia-P. And again, these 51 per cent of iron are redistributed between SNe II and SNe Ia-p depending on input parameters.

       \subsubsection{ {\it Smooth DTD function of Maoz et al. (2010)}}
       
       The radial distribution of iron computed for DTD function of Maoz et al. (2010) is demonstrated in figure 8, the corresponding values, computed for the starting ejected masses, are presented in {\it Case} 6 of Table 2. Comparing the results with the ones of {\it Case} 1 we see that, the constants $\gamma$ and $\zeta$ are changed significantly: $\gamma$ has increased by 2.4 times, $\zeta$ has decreased about 1.6 times. Nevertheless, the masses of iron synthesized for the age of the Galactic disc in the framework of smooth representation for the DTD function happens to be close to the ones corresponding to {\it Case} 1. This statement is also valid for the mass of iron confined in the present ISM.

       \section{Conclusions}

       On the basis of a new observational data on abundances of Cepheids we have studied the problem of how much amount of iron was synthesized by various Types of SNe -- SNe II, prompt and tardy SNe Ia, for the age of the Galactic disc. For this, we develop a statistical method which enables to evaluate the constants $\beta$, $\gamma$ and $\zeta$ for the rates of synthesis of oxygen and iron without any preliminary suppositions like the equipartition among the above 3 Types of SNe. To do that, we develop a theory of iron and oxygen synthesis in the Galactic disc. This theory explains the nontrivial distributions along the Galactic disc of oxygen which demonstrates the bend in the radial gradient at $r \sim 7$ kpc with a rather steep gradient for $5 < r < 7$ kpc and a plateau-like distribution in the region of $7 < r < 10$ kpc, as well as the multi-slope radial distribution of iron in the same range of the Galactocentric radius. In order to understand the mechanism of formation of such fine structure of radial distributions of oxygen and iron we use two main ideas. 
       
       First, there are 2 Types of SNe Ia - {\it prompt} SNe Ia which progenitors are short-lived stars no older than 100 Myr and {\it tardy} SNe Ia whose progenitors may have the ages in the range from 100 Myr to 10 Gyr. For the {\it Delay Time Distribution} function we study both the bimodal approximation of Matteucci et al. (2006) and smooth representation of Maoz et al. (2010).
       
       Second, we take into account the influence of spiral arms on the formation of the fine structure in the radial distribution of oxygen and iron in the Galactic disc. To realize that, we use the representations for the rate of explosions of short-lived SNe progenitors -- SNe II and SNe Ia-P, proposed by Oort (1974), Wyse \& Silk (1989) and Portinari \& Chiosi (1999) (see also Mishurov et al. 2002; Acharova et al. 2005; 2010; 2011). Our statistical method of treatment of the observational data enables to derive simultaneously the location of the corotation resonance which happens to be located at $r_c \approx 7$ kpc and is situated close to the bend in the slope of oxygen distribution or the minimum in $[O/Fe]$. 
       
       Besides, by means of the proposed statistical methods we may estimate the contributions of the 3 Types of SNe to iron synthesis without any preliminary suppositions. The results are as follows. For the age of the Galactic disc about 35 - 38 per cent of iron was produced due to SNe Ia-T and this portion does not varies depending on the input parameters. The total portion of iron produced by SNe II and SNe Ia-P is of the order of 65 per cent. However, the ration of iron between SNe II and SNe Ia-P may changes depending on the ejected mass of oxygen (!) or iron per one SNe II event. Nevertheless, the amounts of iron synthesized by the 3 Types of SNe do not differ significantly from the ones adopted by Matteucci (2004).
       
       However, for the present ISM the situation is another. Thus about 50 per cent of iron in ISM was supplied by SNe Ia-T. The portion of it produced by SNe Ia-P varies from 26 to 42 per cent. Correspondingly, about 9 - 25 per cent of iron, injected by SNe II, was captured by ISM.
       
       At last, the total mass of iron supplied to the Galactic disc during its life by all Types of SNe is $\sim (4.0 \pm 0.4)\cdot 10^7$ M$_{\odot}$, the mass of iron in the present ISM is $\sim (1.20 \pm 0.05)\cdot 10^7 $ M$_{\odot}$ i.e., about 2/3 of iron is contained in stars and stellar remnants. 
       
       Our computations show that the result weakly depend on the exact shape of the DTD function - bimodal (Matteucci et al. 2006) or smooth (Maoz et al. 2010). We only need that there have to be a subpopulation of SNe Ia which progenitors are young, i.e. their ages are not more than 100 Myr in order we can to use the idea that spiral arms influence the formation of radial distribution of iron. Our infer may be considered as an argument in favour of the above estimate for the prompt SNe Ia progenitors.  The result of Bartunov et al. (1994), that a significant portion of SNe Ia is concentrated in spiral arms, supports this idea.

       \section*{Acknowledgments}
                     
       We are gratefull to the anonymos referee for very important comments and suggestions. Authors also thank to Profs. A.Zasov and S.Blinnikov for helpful discussions. The work was supported in part by grants No. 02.740.0247 and P685 of Federal agency for science and innovations. IAA thanks to the Russian funds for basic research, grant No. 11-02-90702. The spectra were collected with the 1.93-m telescope of the OHP (France), the ESO Telescopes at the Paranal Observatory under program ID266.D-5655, and the Mercator Telescope, operated on the island of La Palma by the Flemish Community, at the Spanish Observatorio del Roque de los Muchachos of the Instituto de Astrofisica de Canarias. Drs. C. Soubiran, B. Lemasle, A. Fry and B. Carney are acknowledged for their help with spectral material.


       
       \appendix
       
       
       \setcounter{table}{0}
       \begin{table*}
       \begin{minipage}{14cm}
       \caption{Abundabces, distances, ages and masses for classical Cepheids.}
       \label{tab2}
       \begin{tabular}{rrrrrrrrrrrr}
       \hline
       \hline
            Name &No. Spectra& $P$, days& $V$&$(B-V)$&E({\it B--V}) & r, kpc&  Mv & [O/H]&[Fe/H]&age, Myr &Mass\\
       \hline
            T Ant&   1  &  5.8977098&  9.337& 0.750& 0.300&  8.38  &--3.34 &--0.43&--0.24& 62 & 6.2 \\
            U Aql&   1  &  7.0239582&  6.446& 1.024& 0.399&  7.45  &--3.54 &  0.01&  0.01& 55 & 6.8 \\
           SZ Aql&  11  & 17.1408482&  8.599& 1.389& 0.537&  6.42  &--4.58 &--0.03&  0.17& 30 &10.6 \\
           TT Aql&   8  & 13.7547073&  7.141& 1.292& 0.438&  7.10  &--4.32 &  0.02&  0.10& 35 & 9.5 \\
           FF Aql&  14  &  4.4709158&  5.372& 0.756& 0.196&  7.63  &--3.40 &--0.09&  0.04& 60 & 6.4 \\
           FM Aql&   2  &  6.1142302&  8.270& 1.277& 0.589&  7.29  &--3.38 &--0.19&  0.08& 61 & 6.4 \\
           FN Aql&   4  &  9.4816027&  8.382& 1.214& 0.483&  6.68  &--3.89 &--0.08&--0.02& 45 & 7.9 \\
         V496 Aql&   2  &  6.8070550&  7.751& 1.146& 0.397&  6.88  &--3.89 &--0.15&  0.05& 45 & 7.9 \\
         V600 Aql&   1  &  7.2387481& 10.037& 1.462& 0.798&  6.83  &--3.58 &  0.11&  0.03& 54 & 6.9 \\
         V733 Aql&   1  &  6.1789999&  9.970& 0.960& 0.106&  6.19  &--3.39 &  0.04&  0.08& 60 & 6.4 \\
        V1162 Aql&   2  &  5.3761001&  7.798& 1.366& 0.195&  6.74  &--3.61 &--0.19&  0.01& 53 & 7.0 \\
        V1359 Aql&   1  &  3.7320000&  9.059& 1.350& 0.661&  7.26  &--2.81 &  0.29&  0.09& 84 & 5.0 \\
          Eta Aql&  14  &  7.1767349&  3.897& 0.789& 0.130&  7.71  &--3.57 &--0.06&  0.08& 55 & 6.9 \\
         V340 Ara&   1  & 20.8090000& 10.164& 1.539& 0.546&  4.34  &--4.81 &  0.07&  0.31& 27 &11.7 \\
            Y Aur&   2  &  3.8595021&  9.607& 0.911& 0.375&  9.63  &--2.85 &--0.30&--0.20& 83 & 5.0 \\
           RT Aur&  10  &  3.7281899&  5.446& 0.595& 0.059&  8.30  &--2.81 &--0.07&  0.06& 85 & 5.0 \\
           RX Aur&  16  & 11.6235371&  7.655& 1.009& 0.263&  9.40  &--4.13 &  0.07&--0.01& 39 & 8.8 \\
           SY Aur&   2  & 10.1446981&  9.074& 1.000& 0.432&  9.98  &--3.97 &--0.10&--0.05& 43 & 8.2 \\
           YZ Aur&   5  & 18.1932125& 10.332& 1.375& 0.538& 12.28  &--4.65 &--0.13&--0.35& 29 &11.0 \\
           AN Aur&   4  & 10.2905598& 10.455& 1.218& 0.565& 11.16  &--3.99 &--0.25&--0.15& 43 & 8.2 \\
           AO Aur&   2  &  6.7630062& 10.860& 1.060& 0.431& 11.81  &--3.50 &--0.19&--0.26& 57 & 6.7 \\
           AX Aur&   1  &  3.0466399& 12.412& 1.155& 0.598& 11.95  &--2.57 &--0.02&--0.09& 97 & 4.5 \\
           BK Aur&   2  &  8.0024319&  9.427& 1.062& 0.425& 10.01  &--3.69 &  0.08&  0.06& 51 & 7.3 \\
           CY Aur&   1  & 13.8476496& 11.851& 1.600& 0.768& 13.21  &--4.33 &--0.41&--0.40& 35 & 9.6 \\
           ER Aur&   2  & 15.6907301& 11.520& 1.124& 0.494& 15.36  &--4.48 &--0.63&--0.34& 32 &10.2 \\
         V335 Aur&   1  &  3.4132500& 12.461& 1.137& 0.626& 12.11  &--2.70 &  ... &--0.27& 90 & 4.7 \\
           RW Cam&  16  & 16.4148121&  8.691& 1.351& 0.633&  9.35  &--4.53 &  0.02&  0.09& 31 &10.4 \\
           RX Cam&   9  &  7.9120240&  7.682& 1.193& 0.532&  8.61  &--3.68 &--0.10&  0.04& 51 & 7.2 \\
           TV Cam&   1  &  5.2949700& 11.659& 1.198& 0.613& 11.20  &--3.21 &--0.30&--0.08& 67 & 5.9 \\
           AB Cam&   1  &  5.7876401& 11.849& 1.235& 0.656& 11.44  &--3.32 &--0.29&--0.09& 63 & 6.2 \\
           AD Cam&   1  & 11.2609911& 12.564& 1.588& 0.864& 13.05  &--4.09 &  0.00&--0.22& 40 & 8.6 \\
           RY CMa&   3  &  4.6782498&  8.110& 0.847& 0.239&  8.78  &--3.07 &--0.13&--0.00& 73 & 5.6 \\
           RZ CMa&   3  &  4.2548318&  9.697& 1.004& 0.443&  9.11  &--2.96 &--0.03&--0.03& 77 & 5.3 \\
           TW CMa&   2  &  6.9950700&  9.561& 0.970& 0.329&  9.76  &--3.54 &--0.20&--0.17& 55 & 6.8 \\
           VZ CMa&   1  &  3.1262300&  9.383& 0.957& 0.461&  8.75  &--2.98 &--0.39&--0.06& 76 & 5.4 \\
           AO CMa&   1  &  5.8154202& 12.603& 1.316& 0.738& 11.30  &--3.32 &  ... &--0.14& 63 & 6.2 \\
            U Car&   1  & 38.7681007&  6.288& 1.183& 0.265&  7.54  &--5.53 &  ... &  0.01& 18 &16.0 \\
            V Car&   2  &  6.6966720&  7.362& 0.872& 0.169&  7.88  &--3.49 &--0.15&  0.00& 57 & 6.7 \\
           SX Car&   1  &  4.8600001&  9.089& 0.887& 0.318&  7.59  &--3.11 &--0.30&--0.09& 71 & 5.7 \\
           UW Car&   1  &  5.3457732&  9.426& 0.971& 0.435&  7.62  &--3.22 &--0.28&--0.06& 66 & 5.9 \\
           UX Car&   2  &  3.6822460&  8.308& 0.627& 0.112&  7.66  &--2.79 &--0.05&  0.02& 85 & 4.9 \\
           UY Car&   1  &  5.5437260&  8.967& 0.818& 0.180&  7.55  &--3.27 &--0.15&  0.03& 65 & 6.1 \\
           UZ Car&   1  &  5.2046599&  9.323& 0.875& 0.178&  7.54  &--3.19 &--0.10&  0.07& 68 & 5.9 \\
           VY Car&   1  & 18.9137611&  7.443& 1.171& 0.237&  7.58  &--4.69 &--0.05&  0.12& 28 &11.2 \\
           WW Car&   1  &  4.6768098&  9.743& 0.890& 0.379&  7.52  &--3.07 &--0.55&--0.07& 73 & 5.6 \\
           WZ Car&   1  & 23.0132008&  9.247& 1.142& 0.370&  7.57  &--4.92 &  ... &  0.03& 25 &12.3 \\
           XX Car&   1  & 15.7162399&  9.322& 1.054& 0.347&  7.38  &--4.48 &--0.06&  0.11& 32 &10.2 \\
           XY Car&   1  & 12.4348297&  9.295& 1.214& 0.411&  7.33  &--4.21 &--0.29&  0.04& 38 & 9.1 \\
           XZ Car&   1  & 16.6499004&  8.601& 1.266& 0.365&  7.41  &--4.55 &   ...&  0.14& 31 &10.5 \\
           YZ Car&   1  & 18.1655731&  8.714& 1.124& 0.381&  7.63  &--4.65 &--0.15&  0.02& 29 &11.0 \\
           AQ Car&   2  &  9.7689600&  8.851& 0.928& 0.165&  7.63  &--3.93 &--0.10&  0.00& 44 & 8.0 \\
           CN Car&   1  &  4.9326100& 10.700& 1.089& 0.399&  7.80  &--3.13 &--0.11&  0.06& 70 & 5.7 \\
           CY Car&   1  &  4.2659302&  9.782& 0.953& 0.370&  7.47  &--2.96 &--0.08&  0.10& 77 & 5.3 \\
           DY Car&   1  &  4.6746101& 11.314& 1.003& 0.372&  7.69  &--3.07 &--0.28&--0.07& 73 & 5.6 \\
           ER Car&   2  &  7.7185502&  6.824& 0.867& 0.096&  7.60  &--3.65 &  0.00&  0.01& 52 & 7.1 \\
           FI Car&   1  & 13.4582005& 11.610& 1.514& 0.691&  8.11  &--4.30 &--0.25&  0.06& 36 & 9.4 \\
           FR Car&   1  & 10.7169704&  9.661& 1.121& 0.334&  7.39  &--4.03 &--0.18&  0.02& 42 & 8.4 \\
           GH Car&   1  &  5.7255702&  9.177& 0.932& 0.394&  7.41  &--3.69 &--0.17&--0.01& 51 & 7.2 \\
           GX Car&   1  &  7.1967301&  9.364& 1.043& 0.386&  7.76  &--3.57 &--0.07&  0.01& 54 & 6.9 \\
           HW Car&   1  &  9.2002001&  9.163& 1.055& 0.184&  7.56  &--3.86 &  0.02&  0.04& 46 & 7.8 \\
           IO Car&   1  & 13.5970000& 11.101& 1.221& 0.502&  8.08  &--4.31 &--0.35&--0.05& 36 & 9.5 \\
           IT Car&   1  &  7.5331998&  8.097& 0.990& 0.184&  7.45  &--3.62 &--0.15&  0.06& 53 & 7.1 \\
         V397 Car&   2  &  2.0634999&  8.320& 0.754& 0.266&  7.67  &--2.50 &--0.14&  0.03&101 & 4.4 \\
       \hline
       \hline
       \end{tabular}
       \end{minipage}
       \end{table*}
       
       \setcounter{table}{0}
       \begin{table*}
       \begin{minipage}{14cm}
       \caption{Continued.}
       \label{tab2}
       \begin{tabular}{rrrrrrrrrrrr}
       \hline
       \hline
            Name &No. Spectra& $P$ (days)& V&$(B-V)$&E({\it B--V}) & r, kpc&  Mv & [O/H]&[Fe/H]&age, Myr &Mass\\
       \hline
            l Car &   5  & 35.5513420&  3.724& 1.299& 0.147&  7.79  &--5.43 &  0.12&  0.02& 19 &15.3 \\
           SZ Cas &   1  & 13.6377468&  9.853& 1.419& 0.794&  9.40  &--4.31 &  0.06&  0.04& 35 & 9.5 \\
           RY Cas &   1  & 12.1388798&  9.927& 1.384& 0.618&  9.34  &--4.18 &  0.13&  0.10& 38 & 9.0 \\
           RW Cas &   2  & 14.7915478&  9.117& 1.096& 0.380&  9.96  &--4.41 &--0.07&  0.06& 34 & 9.9 \\
           SU Cas &  13  &  1.9493220&  5.970& 0.703& 0.259&  8.13  &--2.43 &--0.02&  0.06&105 & 4.2 \\
           SW Cas &   1  &  5.4409499&  9.705& 1.081& 0.467&  8.75  &--3.25 &  0.27&  0.02& 66 & 6.0 \\
           SY Cas &   1  &  4.0710979&  9.868& 0.992& 0.442&  8.93  &--2.91 &  0.31&  0.04& 80 & 5.2 \\
           TU Cas &  12  &  2.1392980&  7.733& 0.582& 0.109&  8.31  &--2.16 &--0.03&  0.03&123 & 3.8 \\
           XY Cas &   1  &  4.5016971&  9.935& 1.147& 0.533&  8.98  &--3.02 &--0.09&  0.03& 75 & 5.5 \\
           BD Cas &   3  &  3.6508999& 11.000& 1.565& 1.006&  8.57  &--2.78 &--0.09&--0.07& 86 & 4.9 \\
           CE CasA&   1  &  5.1409001& 10.922& 1.171& 0.556&  9.55  &--3.18 &--0.04&--0.16& 68 & 5.8 \\
           CE CasB&   1  &  4.4793000& 11.062& 1.042& 0.527&  9.62  &--3.02 &--0.04&--0.03& 75 & 5.4 \\
           CF Cas &   5  &  4.8752198& 11.136& 1.174& 0.553&  9.70  &--3.12 &  0.06&--0.01& 71 & 5.7 \\
           CH Cas &   1  & 15.0861902& 10.973& 1.650& 0.894&  9.60  &--4.43 &  ... &--0.08& 33 &10.0 \\
           CY Cas &   1  & 14.3768597& 11.641& 1.738& 0.963& 10.06  &--4.38 &--0.04&  0.06& 34 & 9.7 \\
           DD Cas &   1  &  9.8120270&  9.876& 1.188& 0.450&  9.60  &--3.93 &  0.07&  0.10& 44 & 8.1 \\
           DF Cas &   1  &  3.8324721& 10.848& 1.181& 0.570&  9.72  &--2.84 &  ... &  0.13& 83 & 5.0 \\
           DL Cas &   3  &  8.0006685&  8.969& 1.154& 0.488&  8.85  &--3.69 &--0.01&--0.01& 51 & 7.3 \\
           FM Cas &   1  &  5.8092842&  9.127& 0.989& 0.325&  8.94  &--3.32 &--0.21&--0.09& 63 & 6.2 \\
         V379 Cas &   2  &  4.3057499&  9.053& 1.139& 0.600&  8.59  &--3.36 &  0.07&  0.06& 62 & 6.3 \\
         V636 Cas &   8  &  8.3769999&  7.199& 1.365& 0.666&  8.24  &--3.75 &--0.18&  0.07& 49 & 7.4 \\
            V Cen &   3  &  5.4938612&  6.836& 0.875& 0.292&  7.43  &--3.26 &--0.16&--0.01& 65 & 6.0 \\
           XX Cen &   1  & 10.9533701&  7.818& 0.983& 0.266&  7.00  &--4.06 &--0.03&  0.16& 41 & 8.5 \\
           AY Cen &   1  &  5.3097501&  8.830& 1.009& 0.295&  7.42  &--3.22 &--0.15&  0.01& 67 & 5.9 \\
           AZ Cen &   1  &  3.2119811&  8.636& 0.653& 0.168&  7.41  &--3.01 &--0.10&--0.05& 75 & 5.4 \\
           BB Cen &   1  &  3.9976599& 10.073& 0.953& 0.377&  7.12  &--3.27 &  0.06&  0.13& 65 & 6.1 \\
           KK Cen &   1  & 12.1802998& 11.480& 1.282& 0.611&  7.54  &--4.18 &  0.01&  0.12& 38 & 9.0 \\
           KN Cen &   1  & 34.0296402&  9.870& 1.582& 0.797&  6.40  &--5.38 &  ... &  0.35& 19 &15.0 \\
           MZ Cen &   1  & 10.3529997& 11.531& 1.570& 0.869&  6.53  &--3.99 &--0.13&  0.20& 43 & 8.3 \\
           QY Cen &   1  & 17.7523994& 11.762& 2.150& 1.447&  6.64  &--4.62 &--0.09&  0.16& 30 &10.8 \\
         V339 Cen &   1  &  9.4659996&  8.753& 1.191& 0.413&  6.77  &--3.89 &--0.20&  0.04& 45 & 7.9 \\
         V378 Cen &   1  &  6.4593000&  8.460& 1.035& 0.376&  7.05  &--3.83 &--0.09&--0.02& 47 & 7.7 \\
         V381 Cen &   1  &  5.0787802&  7.653& 0.792& 0.195&  7.24  &--3.17 &--0.09&  0.02& 69 & 5.8 \\
         V419 Cen &   1  &  5.5069098&  8.186& 0.758& 0.168&  7.41  &--3.64 &--0.12&  0.07& 52 & 7.1 \\
         V496 Cen &   1  &  4.4241900&  9.966& 1.172& 0.541&  7.06  &--3.00 &--0.16&  0.00& 75 & 5.4 \\
         V659 Cen &   1  &  5.6217999&  6.598& 0.758& 0.128&  7.45  &--3.28 &  0.00&  0.07& 64 & 6.1 \\
         V737 Cen &   1  &  7.0658498&  6.719& 0.999& 0.206&  7.34  &--3.55 &--0.09&  0.13& 55 & 6.8 \\
           CP Cep &   1  & 17.8589993& 10.590& 1.668& 0.649&  9.60  &--4.63 &--0.16&--0.01& 30 &10.9 \\
           CR Cep &   1  &  6.2329640&  9.656& 1.396& 0.709&  8.44  &--3.79 &--0.09&--0.06& 48 & 7.6 \\
           IR Cep &   2  &  2.1141241&  7.784& 0.888& 0.413&  8.06  &--2.53 &  0.04&  0.05& 99 & 4.4 \\
         V351 Cep &   3  &  2.8060000&  9.440& 0.942& 0.436&  8.49  &--2.86 &  0.07&  0.02& 82 & 5.1 \\
          Del Cep &  18  &  5.3662701&  3.954& 0.657& 0.075&  7.97  &--3.23 &  0.01&  0.09& 66 & 6.0 \\
           AV Cir &   1  &  3.0651000&  7.439& 0.910& 0.378&  7.46  &--2.96 &--0.08&  0.10& 77 & 5.3 \\
           AX Cir &   2  &  5.2733059&  5.880& 0.741& 0.146&  7.53  &--3.21 &--0.07&--0.06& 67 & 5.9 \\
           BP Cir &   1  &  2.3984001&  7.560& 0.702& 0.224&  7.35  &--2.67 &--0.18&--0.06& 91 & 4.7 \\
            R Cru &   1  &  5.8257499&  6.766& 0.772& 0.183&  7.54  &--3.32 &--0.11&  0.08& 63 & 6.2 \\
            S Cru &   1  &  4.6895962&  6.600& 0.761& 0.166&  7.55  &--3.07 &--0.06&--0.12& 73 & 5.6 \\
            T Cru &   1  &  6.7332001&  6.566& 0.922& 0.184&  7.55  &--3.49 &--0.03&  0.09& 57 & 6.7 \\
            X Cru &   1  &  6.2199702&  8.404& 1.001& 0.272&  7.20  &--3.40 &  0.08&  0.14& 60 & 6.4 \\
           VW Cru &   1  &  5.2652202&  9.622& 1.309& 0.643&  7.28  &--3.21 &--0.07&  0.10& 67 & 5.9 \\
           AD Cru &   1  &  6.3978901& 11.051& 1.279& 0.647&  6.99  &--3.43 &--0.06&  0.06& 59 & 6.5 \\
           AG Cru &   1  &  3.8372540&  8.225& 0.738& 0.212&  7.35  &--2.84 &--0.16&--0.13& 83 & 5.0 \\
           BG Cru &   2  &  3.3427200&  5.487& 0.606& 0.132&  7.69  &--3.06 &  0.01&  0.04& 73 & 5.5 \\
            X Cyg &  26  & 16.3863316&  6.391& 1.130& 0.228&  7.73  &--4.53 &  0.07&  0.10& 31 &10.4 \\
           SU Cyg &  12  &  3.8454919&  6.859& 0.575& 0.080&  7.60  &--2.84 &--0.29&--0.03& 83 & 5.0 \\
           SZ Cyg &   1  & 15.1096420&  9.432& 1.477& 0.571&  8.06  &--4.43 &  0.10&  0.09& 33 &10.0 \\
           TX Cyg &   2  & 14.7081566&  9.511& 1.784& 1.130&  7.87  &--4.40 &--0.25&  0.07& 34 & 9.9 \\
           VX Cyg &   1  & 20.1334076& 10.069& 1.704& 0.753&  8.06  &--4.77 &  0.17&  0.09& 27 &11.5 \\
           VY Cyg &   1  &  7.8569822&  9.593& 1.215& 0.606&  7.88  &--3.67 &  0.06&  0.00& 51 & 7.2 \\
           VZ Cyg &   1  &  4.8644528&  8.959& 0.876& 0.266&  8.13  &--3.11 &  0.18&  0.05& 71 & 5.7 \\
           BZ Cyg &   1  & 10.1419315& 10.213& 1.573& 0.888&  7.95  &--3.97 &  ... &  0.07& 43 & 8.2 \\
           CD Cyg &  16  & 17.0739670&  8.947& 1.266& 0.493&  7.47  &--4.58 &--0.03&  0.11& 31 &10.6 \\
           DT Cyg &  14  &  2.4990821&  5.774& 0.538& 0.042&  7.80  &--2.72 &  0.01&  0.10& 89 & 4.8 \\
       \hline
       \hline
       \end{tabular}
       \end{minipage}
       \end{table*}
       
       \setcounter{table}{0}
       \begin{table*}
       \begin{minipage}{14cm}
       \caption{Continued.}
       \label{tab2}
       \begin{tabular}{rrrrrrrrrrrr}
       \hline
       \hline
            Name &No. Spectra& $P$ (days)& V&$(B-V)$&E({\it B--V}) & r, kpc&  Mv & [O/H]&[Fe/H]&age, Myr &Mass\\
       \hline
           MW Cyg &  1  &  5.9545860&  9.489& 1.316& 0.635&  7.55  &--3.35 &  0.14&  0.09& 62 & 6.3 \\
         V386 Cyg &  1  &  5.2576060&  9.635& 1.491& 0.841&  7.89  &--3.21 &--0.06&  0.11& 67 & 5.9 \\
         V402 Cyg &  1  &  4.3648362&  9.873& 1.008& 0.391&  7.60  &--2.99 &  0.11&  0.02& 76 & 5.4 \\
         V532 Cyg &  1  &  3.2836120&  9.086& 1.036& 0.494&  7.98  &--3.04 &  0.03&--0.04& 74 & 5.5 \\
         V924 Cygs&  1  &  5.5714722& 10.710& 0.847& 0.261&  7.53  &--3.65 &  ... &--0.09& 52 & 7.2 \\
        V1154 Cyg &  1  &  4.9254599&  9.190& 0.925& 0.319&  7.70  &--3.13 &  0.00&--0.10& 70 & 5.7 \\
        V1334 Cyg & 11  &  3.3330200&  5.871& 0.504& 0.025&  7.86  &--3.06 &--0.13&  0.03& 73 & 5.5 \\
        V1726 Cyg &  1  &  4.2370601&  9.009& 0.885& 0.339&  8.18  &--3.34 &  ... &--0.02& 62 & 6.2 \\
           TX Dels&  1  &  6.1659999&  9.147& 0.766& 0.222&  6.82  &--3.39 &  0.16&  0.24& 60 & 6.4 \\
         Beta Dor &  1  &  9.8424253&  3.731& 0.807& 0.052&  7.90  &--3.93 &--0.08&--0.01& 44 & 8.1 \\
            W Gem &  8  &  7.9137788&  6.950& 0.889& 0.255&  8.78  &--3.68 &--0.15&--0.01& 51 & 7.2 \\
           RZ Gem &  2  &  5.5292859& 10.007& 1.025& 0.563&  9.84  &--3.26 &--0.14&--0.19& 65 & 6.0 \\
           AA Gem &  2  & 11.3023281&  9.721& 1.061& 0.309& 11.55  &--4.10 &  0.07&--0.27& 40 & 8.6 \\
           AD Gem &  2  &  3.7879801&  9.857& 0.694& 0.173& 10.48  &--2.82 &--0.31&--0.16& 84 & 5.0 \\
           BB Gem &  1  &  2.3080001& 11.364& 0.881& 0.430& 10.64  &--2.25 &--0.45&--0.09&117 & 3.9 \\
           DX Gem &  1  &  3.1374860& 10.746& 0.936& 0.430& 10.76  &--2.99 &--0.28&--0.02& 76 & 5.4 \\
         Zeta Gem & 11  & 10.1507301&  3.918& 0.798& 0.014&  8.25  &--3.97 &--0.05&  0.00& 43 & 8.2 \\
           BB Her &  4  &  7.5079999& 10.090& 1.100& 0.392&  6.05  &--3.62 &  0.04&  0.15& 53 & 7.0 \\
            V Lac &  1  &  4.9834681&  8.936& 0.873& 0.335&  8.48  &--3.14 &  0.17&  0.00& 70 & 5.7 \\
            X Lac &  1  &  5.4449902&  8.407& 0.901& 0.336&  8.48  &--3.63 &  0.10&--0.02& 53 & 7.1 \\
            Y Lac &  9  &  4.3237758&  9.146& 0.731& 0.207&  8.42  &--2.98 &--0.26&--0.04& 77 & 5.3 \\
            Z Lac &  9  & 10.8856134&  8.415& 1.095& 0.370&  8.56  &--4.05 &--0.11&  0.01& 41 & 8.5 \\
           RR Lac &  1  &  6.4162431&  8.848& 0.885& 0.319&  8.55  &--3.44 &  0.09&  0.00& 59 & 6.5 \\
           BG Lac &  3  &  5.3319321&  8.883& 0.949& 0.300&  8.16  &--3.22 &  0.10&--0.01& 67 & 5.9 \\
           GH Lup &  1  &  9.2779484&  7.635& 1.210& 0.335&  6.94  &--3.87 &--0.04&  0.08& 46 & 7.8 \\
         V473 Lyr &  2  &  1.4907800&  6.182& 0.632& 0.025&  7.72  &--2.12 &--0.24&--0.06&125 & 3.7 \\
            T Mon & 20  & 27.0246487&  6.124& 1.166& 0.181&  9.15  &--5.11 &  0.04&  0.13& 22 &13.4 \\
           SV Mon & 10  & 15.2327805&  8.219& 1.048& 0.234& 10.14  &--4.44 &--0.17&--0.02& 33 &10.0 \\
           TW Mon &  2  &  7.0969000& 12.575& 1.339& 0.663& 13.61  &--3.55 &--0.35&--0.18& 55 & 6.8 \\
           TX Mon &  2  &  8.7017307& 10.960& 1.096& 0.485& 11.74  &--3.79 &--0.11&--0.08& 48 & 7.6 \\
           TY Mon &  1  &  4.0226951& 11.740& 1.158& 0.572& 11.07  &--2.89 &  0.01&--0.06& 80 & 5.2 \\
           TZ Mon &  3  &  7.4280138& 10.761& 1.116& 0.420& 11.44  &--3.61 &  0.20&--0.04& 53 & 7.0 \\
           UY Mon &  2  &  2.3979700&  9.391& 0.527& 0.064& 10.08  &--2.67 &--0.16&--0.13& 91 & 4.7 \\
           WW Mon &  2  &  4.6623101& 12.505& 1.128& 0.605& 12.94  &--3.07 &--0.41&--0.36& 73 & 5.6 \\
           XX Mon &  3  &  5.4564729& 11.898& 1.139& 0.567& 11.95  &--3.25 &--0.04&--0.09& 66 & 6.0 \\
           AA Mon &  1  &  3.9381640& 12.707& 1.409& 0.792& 11.36  &--2.87 &--0.19&--0.21& 81 & 5.1 \\
           AC Mon &  2  &  8.0142498& 10.067& 1.165& 0.484& 10.12  &--3.70 &--0.25&--0.22& 51 & 7.3 \\
           BE Mon &  1  &  2.7055099& 10.578& 1.134& 0.622&  9.36  &--2.43 &  0.03&  0.00&105 & 4.2 \\
           BV Mon &  1  &  3.0149601& 11.431& 1.109& 0.612& 10.17  &--2.56 &  ... &--0.14& 97 & 4.5 \\
           CU Mon &  1  &  4.7078729& 13.607& 1.393& 0.751& 14.45  &--3.08 &--0.06&--0.26& 72 & 5.6 \\
           CV Mon &  2  &  5.3788981& 10.299& 1.297& 0.722&  9.46  &--3.23 &--0.12&--0.06& 66 & 6.0 \\
           EE Mon &  1  &  4.8089600& 12.941& 0.966& 0.465& 15.028 &--3.10 &  ... &--0.51& 71 & 5.6 \\
           EK Mon &  2  &  3.9579411& 11.048& 1.195& 0.556& 10.19  &--2.88 &--0.28&--0.06& 81 & 5.1 \\
           FG Mon &  1  &  4.4965901& 13.310& 1.209& 0.651& 13.94  &--3.02 &--0.46&--0.20& 75 & 5.5 \\
           FI Mon &  1  &  3.2878220& 12.924& 1.068& 0.513& 13.11  &--2.66 &--0.41&--0.18& 92 & 4.7 \\
         V465 Mon &  1  &  2.7131760& 10.379& 0.762& 0.244& 10.09  &--2.44 &  0.22&  0.03&105 & 4.2 \\
         V495 Mon &  2  &  4.0965829& 12.427& 1.241& 0.609& 12.10  &--2.92 &--0.09&--0.20& 79 & 5.2 \\
         V504 Mon &  1  &  2.7740500& 11.814& 1.036& 0.538& 10.67  &--2.84 &--0.35&--0.31& 83 & 5.0 \\
         V508 Mon &  2  &  4.1336079& 10.518& 0.898& 0.307& 10.71  &--2.93 &--0.22&--0.21& 79 & 5.2 \\
         V510 Mon &  2  &  7.3071752& 12.681& 1.527& 0.802& 12.96  &--3.59 &--0.20&--0.17& 54 & 6.9 \\
         V526 Mon &  1  &  2.6749849&  8.597& 0.593& 0.089&  9.32  &--2.80 &--0.52&--0.13& 85 & 5.0 \\
            R Mus &  1  &  7.5104671&  6.298& 0.757& 0.114&  7.50  &--3.62 &--0.02&  0.10& 53 & 7.0 \\
            S Mus &  1  &  9.6598749&  6.118& 0.833& 0.212&  7.56  &--3.91 &--0.19&--0.02& 45 & 8.0 \\
           RT Mus &  1  &  3.0861700&  9.022& 0.834& 0.344&  7.43  &--2.59 &  0.02&  0.02& 96 & 4.5 \\
           TZ Mus &  1  &  4.9448848& 11.702& 1.287& 0.664&  7.06  &--3.13 &--0.16&--0.01& 70 & 5.7 \\
           UU Mus &  1  & 11.6364098&  9.781& 1.150& 0.399&  7.05  &--4.13 &--0.06&  0.05& 39 & 8.8 \\
            S Nor &  3  &  9.7542439&  6.394& 0.941& 0.179&  7.17  &--3.92 &--0.13&  0.06& 44 & 8.0 \\
            U Nor &  1  & 12.6437101&  9.238& 1.576& 0.862&  6.82  &--4.23 &  0.04&  0.15& 37 & 9.1 \\
           SY Nor &  1  & 12.6456871&  9.513& 1.340& 0.756&  6.44  &--4.23 &  0.21&  0.31& 37 & 9.1 \\
           TW Nor &  1  & 10.7861795& 11.704& 1.930& 1.157&  5.84  &--4.04 &  0.28&  0.28& 41 & 8.4 \\
           GU Nor &  1  &  3.4528770& 10.411& 1.273& 0.651&  6.55  &--2.72 &  0.15&  0.15& 89 & 4.8 \\
         V340 Nor &  2  & 11.2869997&  8.375& 1.149& 0.321&  6.31  &--4.09 &  0.07&  0.05& 40 & 8.6 \\
            Y Oph & 14  & 17.1269073&  6.169& 1.377& 0.645&  7.42  &--4.58 &  0.00&  0.06& 30 &10.6 \\
       \hline
       \hline
       \end{tabular}
       \end{minipage}
       \end{table*}
       
       \setcounter{table}{0}
       \begin{table*}
       \begin{minipage}{14cm}
       \caption{Continued.}
       \label{tab2}
       \begin{tabular}{rrrrrrrrrrrr}
       \hline
       \hline
            Name &No. Spectra& $P$ (days)& V&$(B-V)$&E({\it B--V}) & r, kpc&  Mv & [O/H]&[Fe/H]&age, Myr &Mass\\
       \hline
           BF Oph &  2  &  4.0675101&  7.337& 0.868& 0.235&  7.12  &--2.91 &--0.08&  0.03& 80 & 5.2 \\
           RS Ori &  6  &  7.5668812&  8.412& 0.945& 0.352&  9.36  &--3.63 &--0.11&--0.06& 53 & 7.1 \\
           CS Ori &  2  &  3.8893900& 11.381& 0.924& 0.383& 11.74  &--2.85 &--0.61&--0.28& 82 & 5.1 \\
           GQ Ori &  2  &  8.6160679&  8.965& 0.976& 0.249& 10.23  &--3.78 &--0.03&  0.01& 48 & 7.5 \\
           SV Per &  1  & 11.1293182&  9.020& 1.029& 0.408& 10.09  &--4.08 &  0.15&  0.01& 41 & 8.6 \\
           UX Per &  1  &  4.5658150& 11.664& 1.027& 0.512& 11.10  &--3.04 &--0.42&--0.21& 74 & 5.5 \\
           VX Per &  9  & 10.8890400&  9.312& 1.158& 0.475&  9.63  &--4.05 &--0.18&--0.04& 41 & 8.5 \\
           AS Per &  1  &  4.9725161&  9.723& 1.302& 0.674&  9.15  &--3.14 &--0.02&  0.10& 70 & 5.7 \\
           AW Per &  4  &  6.4635892&  7.492& 1.055& 0.489&  8.62  &--3.45 &--0.03&  0.01& 58 & 6.5 \\
           BM Per &  4  & 22.9519005& 10.388& 1.793& 0.871& 10.85  &--4.92 &--0.21&  0.00& 25 &12.3 \\
           HQ Per &  2  &  8.6379299& 11.595& 1.234& 0.564& 12.90  &--3.78 &  0.13&--0.31& 48 & 7.6 \\
           MM Per &  1  &  4.1184149& 10.802& 1.062& 0.490& 10.30  &--2.92 &--0.08&--0.01& 79 & 5.2 \\
         V440 Per & 10  &  7.5700002&  6.282& 0.873& 0.260&  8.47  &--3.63 &--0.12&--0.04& 53 & 7.1 \\
            X Pup &  7  & 25.9610004&  8.460& 1.127& 0.402&  9.73  &--5.06 &--0.22&--0.03& 23 &13.1 \\
           RS Pup &  3  & 41.3875999&  6.947& 1.393& 0.457&  8.54  &--5.60 &  0.10&  0.17& 17 &16.5 \\
           VW Pup &  1  &  4.2853699& 11.365& 1.065& 0.489& 10.34  &--2.97 &  0.19&--0.19& 77 & 5.3 \\
           VX Pup &  1  &  3.0108700&  8.328& 0.610& 0.129&  8.64  &--2.56 &--0.05&--0.08& 98 & 4.5 \\
           VZ Pup &  2  & 23.1709995&  9.621& 1.162& 0.459& 10.40  &--4.93 &--0.10&--0.12& 25 &12.4 \\
           WW Pup &  1  &  5.5167241& 10.554& 0.874& 0.379& 10.07  &--3.26 &  ... &--0.18& 65 & 6.0 \\
           WX Pup &  1  &  8.9370499&  9.063& 0.968& 0.319&  9.25  &--3.82 &--0.01&  0.06& 47 & 7.7 \\
           AD Pup &  2  & 13.5939999&  9.863& 1.049& 0.314& 10.61  &--4.31 &  0.03&--0.17& 36 & 9.5 \\
           AP Pup &  5  &  5.0842738&  7.371& 0.838& 0.198&  8.19  &--3.17 &--0.10&  0.00& 69 & 5.8 \\
           AQ Pup &  3  & 30.1040001&  8.791& 1.423& 0.518&  9.49  &--5.23 &  0.04&--0.09& 21 &14.1 \\
           AT Pup &  1  &  6.6648788&  7.957& 0.783& 0.191&  8.41  &--3.48 &--0.31&--0.14& 57 & 6.6 \\
           BC Pup &  2  &  3.5443399& 13.841&    - & 0.800& 12.44  &--2.75 &--0.55&--0.23& 87 & 4.8 \\
           BN Pup &  2  & 13.6731005&  9.882& 1.186& 0.416&  9.92  &--4.32 &  0.00&  0.02& 35 & 9.5 \\
           CE Pup &  1  & 49.5299988& 11.959& 1.745& 0.740& 14.74  &--5.81 &--0.32&--0.05& 15 &18.1 \\
           HW Pup &  3  & 13.4540005& 12.050& 1.237& 0.688& 12.33  &--4.30 &--0.23&--0.23& 36 & 9.4 \\
           MY Pup &  2  &  5.6953092&  5.677& 0.631& 0.061&  8.03  &--3.68 &--0.15&--0.11& 51 & 7.2 \\
           NT Pup &  1  & 15.5649996& 12.144& 1.389& 0.670& 12.42  &--4.47 &--0.39&--0.15& 32 &10.1 \\
         V335 Pup &  1  &  4.8609848&  8.717& 0.759& 0.154&  9.19  &--3.50 &  0.13&--0.01& 57 & 6.7 \\
            S Sge &  9  &  8.3820858&  5.622& 0.805& 0.100&  7.55  &--3.75 &--0.06&  0.08& 49 & 7.4 \\
            U Sgr & 12  &  6.7452288&  6.695& 1.087& 0.403&  7.32  &--3.50 &  0.03&  0.08& 57 & 6.7 \\
            W Sgr &  8  &  7.5949039&  4.668& 0.746& 0.111&  7.51  &--3.63 &--0.14&  0.02& 52 & 7.1 \\
            Y Sgr & 12  &  5.7733798&  5.744& 0.856& 0.191&  7.42  &--3.31 &--0.18&  0.05& 63 & 6.2 \\
           VY Sgr &  1  & 13.5572004& 11.511& 1.941& 1.221&  5.58  &--4.31 &  0.18&  0.26& 36 & 9.5 \\
           WZ Sgr & 12  & 21.8497887&  8.030& 1.392& 0.431&  5.96  &--4.86 &  0.00&  0.19& 26 &12.0 \\
           XX Sgr &  1  &  6.4241400&  8.852& 1.107& 0.521&  6.63  &--3.44 &  0.07&  0.10& 59 & 6.5 \\
           YZ Sgr &  8  &  9.5536871&  7.358& 1.032& 0.281&  6.80  &--3.90 &--0.12&  0.06& 45 & 7.9 \\
           AP Sgr &  1  &  5.0579162&  6.955& 0.807& 0.178&  7.10  &--3.16 &--0.23&  0.10& 69 & 5.8 \\
           AV Sgr &  1  & 15.4150000& 11.391& 1.999& 1.206&  5.48  &--4.46 &  0.36&  0.34& 33 &10.1 \\
           BB Sgr &  1  &  6.6371021&  6.947& 0.987& 0.281&  7.14  &--3.48 &--0.13&  0.08& 57 & 6.6 \\
         V350 Sgr &  1  &  5.1541781&  7.483& 0.905& 0.299&  7.06  &--3.18 &  0.23&  0.18& 68 & 5.8 \\
           RV Sco &  2  &  6.0613060&  7.040& 0.955& 0.349&  7.20  &--3.37 &--0.03&  0.05& 61 & 6.3 \\
           RY Sco &  1  & 20.3201447&  8.004& 1.426& 0.718&  6.67  &--4.78 &  0.06&  0.09& 27 &11.6 \\
           KQ Sco &  1  & 28.6896000&  9.807& 1.934& 0.869&  5.41  &--5.18 &  0.21&  0.16& 22 &13.8 \\
         V482 Sco &  1  &  4.5278072&  7.965& 0.975& 0.336&  6.94  &--3.03 &--0.05&  0.07& 74 & 5.5 \\
         V500 Sco &  5  &  9.3168392&  8.729& 1.276& 0.593&  6.53  &--3.87 &--0.12&  0.01& 46 & 7.8 \\
         V636 Sco &  1  &  6.7968588&  6.654& 0.936& 0.207&  7.15  &--3.50 &--0.08&  0.07& 57 & 6.7 \\
         V950 Sco &  1  &  3.3804500&  7.302& 0.775& 0.254&  7.10  &--3.07 &--0.05&  0.11& 72 & 5.6 \\
            Z Sct &  1  & 12.9013252&  9.600& 1.330& 0.492&  5.52  &--4.25 &  0.16&  0.29& 37 & 9.2 \\
           SS Sct &  1  &  3.6712530&  8.211& 0.944& 0.325&  7.03  &--2.79 &--0.04&  0.06& 85 & 4.9 \\
           UZ Sct &  1  & 14.7441998& 11.305& 1.784& 1.020&  5.12  &--4.40 &  0.49&  0.33& 34 & 9.9 \\
           EV Sct &  1  &  3.0909901& 10.137& 1.160& 0.679&  6.54  &--2.97 &  ... &--0.02& 77 & 5.3 \\
           EW Sct &  3  &  5.8232999&  7.979& 1.725& 1.074&  7.57  &--3.32 &--0.04&  0.04& 63 & 6.2 \\
         V367 Sct &  1  &  6.2930698& 11.596& 1.769& 1.231&  6.43  &--3.41 &  0.53&--0.01& 60 & 6.4 \\
           BQ Ser &  3  &  4.2708998&  9.501& 1.399& 0.815&  7.17  &--2.96 &--0.13&--0.04& 77 & 5.3 \\
           ST Tau &  4  &  4.0342989&  8.217& 0.847& 0.368&  8.83  &--2.90 &--0.12&--0.05& 80 & 5.2 \\
           SZ Tau & 16  &  3.1483800&  6.531& 0.844& 0.295&  8.39  &--2.99 &--0.03&  0.07& 76 & 5.4 \\
           AE Tau &  1  &  3.8964500& 11.679& 1.129& 0.575& 11.33  &--2.86 &--0.17&--0.19& 82 & 5.1 \\
           AV Tau &  1  &  3.6158099& 12.338& 1.376& 0.892& 10.67  &--2.77 &  ... &--0.09& 86 & 4.9 \\
           EF Tau &  1  &  3.4481499& 13.113& 0.931& 0.360& 16.32  &--2.71 &--0.23&--0.74& 89 & 4.8 \\
           EU Tau &  2  &  2.1024799&  8.093& 0.664& 0.164&  8.93  &--2.52 &--0.05&--0.06&100 & 4.4 \\
       \hline
       \hline
       \end{tabular}
       \end{minipage}

       \end{table*}
       
       \setcounter{table}{0}
       \begin{table*}
       \begin{minipage}{14cm}
       \caption{Continued.}
       \label{tab2}
       \begin{tabular}{rrrrrrrrrrrr}
       \hline
       \hline
            Name &No. Spectra& $P$ (days)& V&$(B-V)$&E({\it B--V}) & r, kpc&  Mv & [O/H]&[Fe/H]&age, Myr &Mass\\
       \hline
            R TrA &  1  &  3.3892870&  6.660& 0.722& 0.142&  7.48  &--2.69 &--0.08&  0.06& 90 & 4.7 \\
            S TrA &  1  &  6.3234649&  6.397& 0.752& 0.084&  7.28  &--3.42 &--0.13&  0.12& 59 & 6.5 \\
           LR TrA &  1  &  2.4549999&  7.808& 0.781& 0.268&  7.29  &--2.70 &  0.03&  0.25& 90 & 4.7 \\
          Alp UMi &  1  &  3.9696000&  1.982& 0.598& 0.000&  7.96  &--3.26 &  0.12&  0.10& 65 & 6.0 \\
            T Vel &  4  &  4.6398191&  8.024& 0.922& 0.289&  8.05  &--3.06 &--0.05&--0.02& 73 & 5.5 \\
            V Vel &  2  &  4.3710432&  7.589& 0.788& 0.186&  7.85  &--2.99 &--0.25&--0.21& 76 & 5.4 \\
           RY Vel &  4  & 28.1357002&  8.397& 1.352& 0.547&  7.73  &--5.16 &--0.03&  0.05& 22 &13.6 \\
           RZ Vel &  4  & 20.3982391&  7.079& 1.120& 0.299&  8.22  &--4.78 &--0.03&  0.04& 27 &11.6 \\
           ST Vel &  2  &  5.8584251&  9.704& 1.195& 0.479&  8.18  &--3.33 &--0.26&  0.02& 62 & 6.2 \\
           SV Vel &  1  & 14.0970697&  8.524& 1.054& 0.373&  7.59  &--4.35 &--0.16&  0.08& 35 & 9.7 \\
           SW Vel &  5  & 23.4410000&  8.120& 1.162& 0.344&  8.43  &--4.94 &--0.11&--0.10& 25 &12.4 \\
           SX Vel &  4  &  9.5499296&  8.251& 0.888& 0.263&  8.24  &--3.90 &--0.03&--0.02& 45 & 7.9 \\
           XX Vel &  1  &  6.9845700& 10.654& 1.162& 0.545&  7.71  &--3.54 &--0.29&--0.05& 56 & 6.8 \\
           AE Vel &  1  &  7.1335702& 10.262& 1.243& 0.635&  7.98  &--3.56 &--0.03&  0.05& 55 & 6.9 \\
           AH Vel &  3  &  4.2272310&  5.695& 0.579& 0.070&  8.00  &--3.33 &  0.00&  0.05& 62 & 6.2 \\
           AX Vel &  1  &  3.6731000&  8.197& 0.691& 0.224&  8.11  &--2.79 &  ... &--0.08& 85 & 4.9 \\
           BG Vel &  2  &  6.9236550&  7.635& 1.175& 0.426&  7.92  &--3.53 &  0.01&--0.02& 56 & 6.8 \\
           CS Vel &  1  &  5.9047399& 11.681& 1.448& 0.737&  8.20  &--3.34 &--0.01&  0.08& 62 & 6.2 \\
           CX Vel &  1  &  6.2554250& 11.374& 1.413& 0.723&  8.36  &--3.41 &--0.30&  0.06& 60 & 6.4 \\
           DK Vel &  1  &  2.4816401& 10.614& 0.774& 0.287&  8.13  &--2.33 &  0.03&--0.02&111 & 4.0 \\
           DR Vel &  2  & 11.1992998&  9.520& 1.518& 0.656&  8.04  &--4.08 &--0.02&  0.08& 40 & 8.6 \\
           EX Vel &  1  & 13.2341003& 11.562& 1.561& 0.775&  8.87  &--4.28 &--0.11&  0.05& 36 & 9.4 \\
           EZ Vel &  2  & 34.5345993& 12.448& 1.716& 0.822& 12.51  &--5.39 &--0.01&--0.08& 19 &15.1 \\
           FG Vel &  1  &  6.4531999& 11.814& 1.493& 0.810&  8.29  &--3.44 &--0.06&--0.05& 59 & 6.5 \\
           FN Vel &  1  &  5.3242202& 10.292& 1.186& 0.588&  7.85  &--3.22 &--0.17&  0.06& 67 & 5.9 \\
            S Vul &  4  & 68.4639969&  8.962& 1.892& 0.727&  7.07  &--6.19 &--0.20&--0.01& 12 &21.3 \\
            T Vul & 12  &  4.4354620&  5.754& 0.635& 0.064&  7.76  &--3.01 &--0.09&  0.01& 75 & 5.4 \\
            U Vul &  7  &  7.9906292&  7.128& 1.275& 0.603&  7.58  &--3.69 &--0.04&  0.09& 51 & 7.3 \\
            X Vul &  6  &  6.3195429&  8.849& 1.389& 0.742&  7.53  &--3.42 &--0.03&  0.07& 59 & 6.5 \\
           SV Vul & 23  & 44.9947739&  7.220& 1.442& 0.461&  7.26  &--5.70 &--0.01&  0.05& 16 &17.2 \\
       \hline
       \hline
       \end{tabular}
       \end{minipage}
       \\
       \end{table*}

       \bsp
                     
              \label{lastpage}
                     
              \end{document}